\documentstyle[12pt]{article}

\begin{document}

\begin{titlepage}
\begin{flushright}CERN-TH/98-360\end{flushright}$ $\vspace{1cm}\\
\centerline{\bf Duality symmetry of Reggeon interactions in
multicolour QCD}
\date{}
\begin{center}
L.N. Lipatov {$^{*}$\\
Petersburg Nuclear Physics Institute,\\
Gatchina, 188350, St.Petersburg, Russia}
\end{center}
\vskip 15.0pt
\centerline{\bf Abstract}
\noindent
The duality symmetry of the Hamiltonian and integrals of
motion for Reggeon interactions in multicolour QCD is
formulated as an integral equation for the wave function of
compound states of $n$ reggeized gluons. In particular
the Odderon problem in QCD is reduced to the solution of the
one-dimensional Schr\"odinger equation. The Odderon
Hamiltonian is written in a normal form, which gives a
possibility to express it as a function of its integrals of motion.
\vskip 3cm \hrule
\vskip 3cm \noindent
\noindent $^{*}$ {\it Supported by the CRDF, INTAS and INTAS-RFBR grants:
RP1-253, 1867-93, 95-0311}
\vfill
\end{titlepage}

\section{Introduction}

The hadron scattering amplitude at high energies $\sqrt{s}$ in the leading
logarithmic approximation (LLA) of the perturbation theory is obtained by
calculating and summing all contributions $\left( g^{2}\ln (s)\right) ^{n}$,
where $g$ is the coupling constant. In this approximation the gluon is
reggeized and the BFKL Pomeron is a compound state of two reggeized gluons
[1]. Next-to-leading corrections to the BFKL equation were also calculated
[2], which gives a possibility to find its region of applicability. In
particular the M\"{o}bius invariance of the equation valid in LLA [1] turns
out to be violated after taking into account next-to-leading terms.

The asymptotic behaviour $\propto s^{j_{0}}$ of scattering amplitudes is
governed by the $j$-plane singularities of the $t$-channel partial waves $%
f_{j}(t)$. The position of these singularities $\omega _{0}=j_{0}-1$ for the
Feynman diagrams with $n$ reggeized gluons in the $t$-channel is
proportional to eigenvalues of a Schr\"{o}dinger-like equation [3]. For the
multicolour QCD $N_{c}\rightarrow \infty $ the colour structure and the
coordinate dependence of the eigenfunctions are factorized [4].

The wave function $f_{m,\widetilde{m}}(\overrightarrow{\rho _{1}},%
\overrightarrow{\rho _{2}},...,\overrightarrow{\rho _{n}};\overrightarrow{%
\rho _{0}})$ of the colourless compound state $O_{m\widetilde{,m}}(%
\overrightarrow{\rho _{0}})$ depends on the two-dimensional impact
parameters $\overrightarrow{\rho _{1}},\overrightarrow{\rho _{2}},...,%
\overrightarrow{\rho _{n}}$ of the reggeized gluons. It belongs to the basic
series of unitary representations of the M\"{o}bius group transformations

\begin{equation}
\rho _{k}\rightarrow \frac{a\,\rho _{k}+b}{c\,\rho _{k}+d}\,,
\end{equation}
where $\rho _{k}=x_{k}+iy_{k},\,\rho _{k}^{*}=x_{k}-iy_{k}$ and $a,b,c,d$
are arbitrary complex parameters [1]. For this series the conformal weights

\begin{equation}
m=1/2+i\nu +n/2,\,\,\widetilde{m}=1/2+i\nu -n/2
\end{equation}
are expressed in terms of the anomalous dimension $\gamma =1+2i\nu $ and the
integer conformal spin $n$ of the composite operators $O_{m\widetilde{,m}}(%
\overrightarrow{\rho _{0}})$. They are related to the eigenvalues

\begin{equation}
M^{2}f_{m,\widetilde{m}}=m(m-1)f_{m,\widetilde{m}}\,,\,\,\,M^{*2}f_{m,%
\widetilde{m}}=\widetilde{m}(\widetilde{m}-1)f_{m,\widetilde{m}}\,
\end{equation}
of the Casimir operators $M^{2}$ and $M^{*2}$:
\begin{equation}
M^{2}=\left( \sum_{k=1}^{n}M_{k}^{a}\right)
^{2}=\sum_{r<s}2\,M_{r}^{a}\,M_{s}^{a}=-\sum_{r<s}\rho _{rs}^{2}\partial
_{r}\partial _{s}\,,\,\, M^{*2}=(M^{2})^{*}.
\end{equation}
Here $M_{k}^{a}$ are the M\"{o}bius group generators
\begin{equation}
M_{k}^{3}=\rho _{k}\partial _{k}\,,\,\,\,M_{k}^{-}=\partial
_{k}\,,\,\,\,M_{k}^{+}=-\rho _{k}^{2}\partial _{k}
\end{equation}
and $\partial _{k}=\partial /(\partial \rho _{k})$.

The wave function $f_{m,\widetilde{m}}$ satisfies the Schr\"{o}dinger
equation [4]:

\begin{equation}
E_{m,\widetilde{m}}\,f_{m,\widetilde{m}}=H\,f_{m,\widetilde{m}}.
\end{equation}
Its eigenvalue $E_{m,\widetilde{m}}$ is proportional to the position $\omega
_{m,\widetilde{m}}=j-1$ of a $j$-plane singularity of the $t$-channel
partial wave:

\begin{equation}
\omega _{m,\widetilde{m}}\,=-\frac{g^{2}N_{c}}{8\,\pi ^{2}}\,E_{m,\widetilde{%
m}}
\end{equation}
governing the $n$-Reggeon asymptotic contribution to the total cross-section
$\sigma _{tot}\sim s^{\omega _{m,\widetilde{m}}}$.

In the particular case of the Odderon, being a compound state of three
reggeized gluons with the charge parity $C=-1$ and the signature $P_{j}=-1$,
the eigenvalue $\omega _{m,\widetilde{m}}^{(3)}$ is related to the
high-energy behaviour of the difference of the total cross-sections $\sigma
_{pp}$ and $\sigma _{p\overline{p}}$ for interactions of particles $p$ and
antiparticles $\overline{p}$ with a target:

\begin{equation}
\sigma _{pp}-\sigma _{p\overline{p}}\sim s^{\omega _{m,\widetilde{m}}^{(3)}}.
\end{equation}

The Hamiltonian $H$ in the multicolour QCD has the property of the
holomorphic separability [4]:

\begin{equation}
H=\frac{1}{2}(h+h^{*}),\,\,\,\left[ h,h^{*}\right] =0\,,
\end{equation}
where the holomorphic and anti-holomorphic Hamiltonians

\begin{equation}
h=\sum_{k=1}^{n}h_{k,k+1\,},\,h^{*}=\sum_{k=1}^{n}h_{k,k+1\,}^{*}
\end{equation}
are expressed in terms of the BFKL operator [4] :

\begin{equation}
h_{k,k+1}=\log (p_{k})+\log (p_{k+1})+\frac{1}{p_{k}}\log (\rho
_{k,k+1})p_{k}+\frac{1}{p_{k+1}}\log (\rho _{k,k+1})p_{k+1}+2\,\gamma \,.
\end{equation}
Here $\rho _{k,k+1}=\rho _{k}-\rho _{k+1}\,,\,p_{k}=i\,\partial /(\partial
\rho _{k}),\,p_{k}^{*}=i\,\partial /(\partial \rho _{k}^{*})\,$, and $\gamma
=-\psi (1)$ is the Euler constant.

Owing to the holomorphic separability of $h$, the wave function $f_{m,%
\widetilde{m}}(\overrightarrow{\rho _{1}},\overrightarrow{\rho _{2}},...,%
\overrightarrow{\rho _{n}};\overrightarrow{\rho _{0}})$ has the property of
the holomorphic factorization [4]:

\begin{equation}
f_{m,\widetilde{m}}(\overrightarrow{\rho _{1}},\overrightarrow{\rho _{2}}%
,...,\overrightarrow{\rho _{n}};\overrightarrow{\rho _{0}}%
)=\sum_{r,l}c_{r,l}\,f_{m}^{r}(\rho _{1},\rho _{2},...,\rho _{n};\rho
_{0})\,f_{\widetilde{m}}^{l}(\rho _{1}^{*},\rho _{2}^{*},...,\rho
_{n}^{*};\rho _{0}^{*})\,,
\end{equation}
where $r$ and $l$ enumerate degenerate solutions of the Schr\"{o}dinger
equations in the holomorphic and anti-holomorphic sub-spaces:

\begin{equation}
\epsilon _{m}\,f_{m}=h\,f_{m}\,,\,\,\,\epsilon _{\widetilde{m}}\,f_{%
\widetilde{m}}=h^{*}\,f_{\widetilde{m}}\,,\,\,\,E_{m,\widetilde{m}}=\epsilon
_{m}+\epsilon _{\widetilde{m}}\,.
\end{equation}
Similarly to the case of two-dimensional conformal field theories, the
coefficients $c_{r,l}$ are fixed by the single-valuedness condition for the
function $f_{m,\widetilde{m}}(\overrightarrow{\rho _{1}},\overrightarrow{%
\rho _{2}},...,\overrightarrow{\rho _{n}};\overrightarrow{\rho _{0}})$ in
the two-dimensional $\overrightarrow{\rho }$-space.

There are two different normalization conditions for the wave function
[4],[5]:

\begin{equation}
\left\| f\right\| _{1}^{2}=\int \prod_{r=1}^{n}d^{2}\rho _{r}\,\left|
\prod_{r=1}^{n}\rho _{r,r+1}^{-1}\,\,f\right| ^{2}\,,\,\,\,\left\| f\right\|
_{2}^{2}=\int \prod_{r=1}^{n}d^{2}\rho _{r}\left|
\prod_{r=1}^{n}p_{r}\,\,f\right| ^{2}
\end{equation}
compatible with the hermicity properties of $H$. Indeed, the transposed
Hamiltonian $h^{t}$ is related with $h$ by two different similarity
transformations [5]:

\begin{equation}
h^{t}=\prod_{r=1}^{n}p_{r}\,h\,\prod_{r=1}^{n}p_{r}^{-1}=\prod_{r=1}^{n}\rho
_{r,r+1}^{-1}\,h\,\prod_{r=1}^{n}\rho _{r,r+1.}
\end{equation}

Therefore $h$ commutes

\begin{equation}
\left[ h,A\right] =0
\end{equation}
with the differential operator [4]

\begin{equation}
A=\rho _{12}\rho _{23}...\rho _{n1}\,p_1p_2...p_n\,.
\end{equation}

Furthermore [5], there is a family $\{q_{r}\}$\thinspace of mutually
commuting integrals of motion:

\begin{equation}
\left[ q_{r},q_{s}\right] =0\,,\,\,\,\left[ q_{r},h\right] =0.
\end{equation}
They are given as

\begin{equation}
q_{r}=\sum_{i_{1}<i_{2}<...<i_{r}}\rho _{i_{1}i_{2}\,}\rho
_{i_{2}i_{3}}...\rho _{i_{r}i_{1}}\,p_{i_{1}}\,p_{i_{2}}...p_{i_{r}}.
\end{equation}
In particular $q_{n}$ is equal to $A$ and $q_{2}$ is proportional to $M^{2}$.

The generating function for these integrals of motion coincides with the
transfer matrix $T$ for the $XXX$ model [5]:

\begin{equation}
T(u)=tr\,(L_{1}(u)L_{2}(u)...L_{n}(u))=\sum_{r=0}^{n}u^{n-r}\,q_{r},
\end{equation}
where the $L$-operators are

\begin{equation}
L_{k}(u)=\left( 
\begin{array}{cc}
u+\rho _{k}\,p_{k} & p_{k} \\ 
-\rho _{k}^{2}\,p_{k} & u-\rho _{k}\,p_{k}
\end{array}
\right) =u\,\left( 
\begin{array}{cc}
1 & 0 \\ 
0 & 1
\end{array}
\right) +\left( 
\begin{array}{c}
1 \\ 
-\rho _{k}
\end{array}
\right) \,\left( 
\begin{array}{cc}
\rho _{k} & 1
\end{array}
\right) \,p_{k}\,.
\end{equation}
The transfer matrix is the trace of the monodromy matrix $t(u)$ [6]:

\begin{equation}
T(u)=tr\,(t(u)),\,\,t(u)=L_{1}(u)L_{2}(u)...L_{n}(u)\,.
\end{equation}

It can be verified that $t(u)$ satisfies the Yang-Baxter equation [5],[6]:

\begin{equation}
t_{r_{1}^{\prime }}^{s_{1}}(u)\,t_{r_{2}^{\prime
}}^{s_{2}}(v)\,l_{r_{1}r_{2}}^{r_{1}^{\prime }r_{2}^{\prime
}}(v-u)=l_{s_{1}^{\prime }s_{2}^{\prime
}}^{s_{1}s_{2}}(v-u)\,t_{r_{2}}^{s_{2}^{\prime
}}(v)\,t_{r_{1}}^{s_{1}^{\prime }}(u)\,,
\end{equation}
where $l(w)$ is the $L$-operator for the well-known Heisenberg spin model:

\begin{equation}
l_{s_{1}^{\prime }s_{2}^{\prime }}^{s_{1}s_{2}}(w)=w\,\delta _{s_{1}^{\prime
}}^{s_{1}}\,\delta _{s_{2}^{\prime }}^{s_{2}}+i\,\delta _{s_{2}^{\prime
}}^{s_{1}}\,\delta _{s_{1}^{\prime }}^{s_{2}}\,.
\end{equation}
The commutativity of $T(u)$ and $T(v)$ 
\begin{equation}
\lbrack T(u),T(v)]=0
\end{equation}
is a consequence of the Yang-Baxter equation.

If one will parametrize $t(u)$ in the form

\begin{equation}
t(u)=\left( 
\begin{array}{cc}
j_{0}(u)+j_{3}(u) & j_{-}(u) \\ 
j_{+}(u) & j_{0}(u)-j_{3}(u)
\end{array}
\right) ,
\end{equation}
this equation is reduced to the following Lorentz-covariant relations for
the currents $j_{\mu }(u)$:

\begin{equation}
\left[ j_{\mu }(u),j_{\nu }(v)\right] =\left[ j_{\mu }(v),j_{\nu }(u)\right]
=\frac{i\,\epsilon _{\mu \nu \rho \sigma }}{2(u-v)}\left( j^{\rho
}(u)j^{\sigma }(v)-j^{\rho }(v)j^{\sigma }(u)\right) .
\end{equation}
Here $\epsilon _{\mu \nu \rho \sigma }$ is the antisymmetric tensor ($%
\epsilon _{1230}=1$) in the four-dimensional Minkovski space and the metric
tensor $g^{\mu \nu }$ has the signature ($1,-1,-1,-1$). This form follows
from the invariance of the Yang-Baxter equations under the Lorentz
transformations.

The generators for the spacial rotations coincide with that of the
M\"{o}bius transformations

\begin{equation}
\overrightarrow{M}=\sum_{k=1}^n\overrightarrow{M_k}\,,
\end{equation}

\begin{equation}
M_{k}^{3}=\rho _{k}\,\partial _{k}\,,\,\,M_{k}^{1}=\frac{1}{2}\,(1-\rho
_{k}^{2})\,\partial _{k}\,,\,\,M_{k}^{2}=\frac{i}{2}\,(1+\rho
_{k}^{2})\,\partial _{k}\,\,.
\end{equation}
The commutation relations for the Lorentz algebra are given below:

\begin{equation}
\left[ M^{s},M^{t}\right] =i\epsilon _{stu}\,M^{u},\,\,\,\left[
M^{s},N^{t}\right] =i\epsilon _{stu}\,N^{u},\,\,\,\left[ N^{s},N^{t}\right]
=i\epsilon _{stu}\,M^{u}\,,
\end{equation}
where $\overrightarrow{N}$ are the Lorentz boost generators.

The commutativity of the transfer matrix $T(u)$ with the local hamiltonian $%
h $ [5],[7]

\begin{equation}
\left[ T(u),h\right] =0
\end{equation}
is a consequence of the relation:

\begin{equation}
\left[ L_{k}(u)\,L_{k+1}(u),h_{k,k+1}\right] =-i\,(L_{k}(u)-L_{k+1}(u))
\end{equation}
for the pair Hamiltonian $h_{k,k+1}$. In turn, this relation follows from
the M\"{o}bius invariance of $h_{k,k+1}$ and the identity:

\begin{equation}
\left[ h_{k,k+1},\left[ \left( \overrightarrow{M_{k,k+1}}\right) ^{2},%
\overrightarrow{N_{k,k+1}}\right] \right] =4\,\overrightarrow{N_{k,k+1}}\,,
\end{equation}
where

\begin{equation}
\overrightarrow{M_{k,k+1}}=\overrightarrow{M_{k}}+\,\overrightarrow{M_{k+1}}%
\,,\,\,\,\,\,\overrightarrow{N_{k,k+1}}=\overrightarrow{M_{k}}-\,%
\overrightarrow{M_{k+1}}
\end{equation}
are the Lorentz group generators for the two gluon state.

Because the pair hamiltonian $h_{k,k+1}$ depends only on the Casimir
operator $\left( \overrightarrow{M_{k,k+1}}\right) ^{2}$, it is diagonal

\begin{equation}
h_{k,k+1}\left| m_{k,k+1}\right\rangle =\left( \psi (m_{k,k+1})+\psi
(1-m_{k,k+1})-2\,\psi (1)\right) \left| m_{k,k+1}\right\rangle
\end{equation}
in the conformal weight representation:

\begin{equation}
\left( \overrightarrow{M_{k,k+1}}\right) ^2\left| m_{k,k+1}\right\rangle
=m_{k,k+1}(m_{k,k+1}-1)\left| m_{k,k+1}\right\rangle \,.
\end{equation}

Using the commutation relations between $\overrightarrow{M_{k,k+1}}$ and $%
\overrightarrow{N_{k,k+1}}$ and taking into account that $\left( 
\overrightarrow{M_{k}}\right) ^{2}=0$, one can verify that the operator $%
\overrightarrow{N_{k,k+1}}$ has non-vanishing matrix elements only between
the states $\left| m_{k,k+1}\right\rangle $ and $\left| m_{k,k+1}\pm
1\right\rangle $. It means that the above identity for $h_{k,k+1}$ is a
consequence of the known recurrence relations for the $\psi $-functions:

\begin{equation}
\psi (m)=\psi (m-1)+1/(m-1)\,,\,\,\,\,\psi (1-m)=\psi (2-m)+1/(m-1).
\end{equation}

The pair Hamiltonian $h_{k,k+1}$ can be expressed in terms of the small-$u$
asymptotics

\begin{equation}
\widehat{L}_{k,k+1}(u)=P_{k,k+1}(1+i\,u\,h_{k,k+1}+...)
\end{equation}
of the fundamental $L$-operator $\widehat{L}_{k,k+1}(u)$ acting on functions 
$f(\rho _{k},\rho _{k+1})$ [6]. Here $P_{k,k+1}$ is defined by the relation:

\begin{equation}
P_{k,k+1}\,\,f(\rho _{k},\rho _{k+1})=f(\rho _{k+1},\rho _{k}).
\end{equation}

The operator $\widehat{L}_{k,k+1}$ satisfies the linear equation [6]

\begin{equation}
L_{k}(u)\,L_{k+1}(v)\,\,\widehat{L}_{k,k+1}(u-v)=\widehat{L}%
_{k,k+1}(u-v)\,\,L_{k+1}(v)\,L_{k}(u).
\end{equation}
This equation can be solved in a way similar to that for $h_{k,k+1}$%
\begin{equation}
\widehat{L}_{k,k+1}(u)\sim \,P_{k,k+1}\sqrt{\frac{\Gamma (\widehat{m}%
_{k,k+1}+iu)\Gamma (1-\widehat{m}_{k,k+1}+iu)}{\Gamma (\widehat{m}%
_{k,k+1}-iu)\Gamma (1-\widehat{m}_{k,k+1}-iu)}}\frac{\Gamma (1-iu)}{\Gamma
(1+iu)}\,,
\end{equation}
where the integral operator $\widehat{m}_{k,k+1}$ is defined by the relation

\begin{equation}
\widehat{m}_{k,k+1}\,(\widehat{m}_{k,k+1}-1)=\left( \overrightarrow{M_{k}}+%
\overrightarrow{M_{k+1}}\right) ^{2}
\end{equation}
and the proportionality constant, being a periodic function of $\widehat{m}%
_{k,k+1}$ with an unit period, is fixed from the triangle relation

\begin{equation}
\widehat{L}_{13}(u)\,\widehat{L}_{23}(v)\,\,\widehat{L}_{12}(u-v)=\widehat{L}%
_{12}(u-v)\,\,\widehat{L}_{23}(v)\,\widehat{L}_{13}(u)\,.
\end{equation}

To find a representation of the Yang-Baxter commutation relations, the
algebraic Bethe anzatz is used [6]. To begin with, in the above
parametrization of the monodromy matrix $t(u)$ in terms of the currents $%
j_{\mu }(u)$, one should construct the pseudovacuum state $|0\rangle $
satisfying the equations

\begin{equation}
j_{+}(u)\,|0\rangle =0.
\end{equation}
However, these equations have a non-trivial solution only if the above $L$%
-operators are regularized as

\begin{equation}
L_{k}^{\delta }(u)=\left( 
\begin{array}{cc}
u+\rho _{k}\,p_{k}-i\,\delta & p_{k} \\ 
-\rho _{k}^{2}\,p_{k}+2\,i\,\rho _{k}\,\delta & u-\rho _{k}\,p_{k}+i\delta
\end{array}
\right)
\end{equation}
by introducing an infinitesimally small conformal weight $\delta \,\,$$%
\rightarrow 0$ for reggeized gluons (another possibility is to use the dual
space corresponding to $\delta =-1$ [7]). For this regularization the
pseudovacuum state is

\begin{equation}
|\delta \rangle =\prod_{k=1}^{n}\rho _{k}^{2\delta }
\end{equation}
It is also an eigenstate of the transfer matrix:

\begin{equation}
T(u)\,|\delta \rangle =2\,j_{0}(u)\,|\delta \rangle =\left( (u-i\,\delta
)^{n}+(u+i\,\delta )^{n}\right) \,|\delta \rangle .
\end{equation}

Furthermore, all excited states are obtained by applying the product of the
operators $j_{-}(v)$ to the pseudovacuum state:

\begin{equation}
|v_{1}v_{2}...v_{k}\rangle
=j_{-}(v_{1})\,j_{-}(v_{2})...j_{-}(v_{k})\,\,|\delta \rangle .
\end{equation}
They are eigenfunctions of the transfer matrix $T(u)$ with the eigenvalues:

\begin{equation}
\widetilde{T}(u)=(u+i\delta )^{n}\prod_{r=1}^{k}\frac{u-v_{r}-i}{u-v_{r}}%
+(u-i\delta )^{n}\prod_{r=1}^{k}\frac{u-v_{r}+i}{u-v_{r}}\,,
\end{equation}
providing that the spectral parameters $v_{1},v_{2},...,v_{k}$ are solutions
of the set of Bethe equations:

\begin{equation}
\left( \frac{v_{s}-i\delta }{v_{s}+i\delta }\right) ^{n}=\prod_{r\neq s}%
\frac{v_{s}-v_{r}-i}{v_{s}-v_{r}+i}
\end{equation}
for $s=1,2...k$.

Due to above relations the function

\begin{equation}
Q(u)=\prod_{r=1}^{k}(u-v_{r})\,
\end{equation}
satisfies the Baxter equation [6,7]:

\begin{equation}
\widetilde{T}(u)\,Q(u)=(u-i\delta )^{n}\,Q(u+i)+(u+i\delta )^{n}\,Q(u-i)\,,
\end{equation}
where $\widetilde{T}(u)$ is an eigenvalue of the transfer matrix. Its
corresponding eigenfunctions can be expressed in terms of the solution $%
Q^{(k)}(u)$ of this equation as follows [7]

\begin{equation}
|v_{1}v_{2}...v_{k}\rangle =Q^{(k)}(\widehat{u}_{1})\,\,Q^{(k)}(\widehat{u}%
_{2})\,...\,Q^{(k)}(\widehat{u}_{n-1})\,|\delta \rangle \,,
\end{equation}
where the integral operators $\widehat{u}_{r}$ are zeros of the current $%
j_{-}(u)$: 
\begin{equation}
j_{-}(u)=c\prod_{r=1}^{n-1}(u-\widehat{u}_{r})\,.
\end{equation}

Eigenvalues $\epsilon $ of the holomorphic Hamiltonian $h$ also can be
expressed in terms of $Q(u)$ [7]. Up to now the Baxter equation was solved
only for the case of the BFKL Pomeron ($n=2$). This is the reason why we use
below another approach, based on the diagonalization of the transfer matrix.

\section{Duality of Reggeon interactions at large N$_{c}$}

The differential operators $q_{r}$ and the Hamiltonian $h$ are invariant
under the cyclic permutation of gluon indices $i\rightarrow i+1$ ($i=1,2...n$%
), corresponding to the Bose symmetry of the Reggeon wave function at $%
N_{c}\rightarrow \infty $. It is remarkable that above operators are
invariant also under the more general canonical transformation:

\begin{equation}
\rho _{i-1,i}\rightarrow p_{i}\rightarrow \rho _{i,i+1}\,,\,
\end{equation}
combined with reversing the order of the operator multiplication.

This invariance is obvious for the Hamiltonian $h$ if we write it in the
form:

\begin{equation}
h=h_{p}+h_{\rho }\,,
\end{equation}
where 
\[
h_{p}=\sum_{k=1}^{n}\left( \ln (p_{k})+\frac{1}{2}\sum_{\lambda =\pm 1} \rho
_{k,k+\lambda }\,\ln (p_{k})\,\rho _{k,k+\lambda }^{-1}+\gamma \right) \,, 
\]
\begin{equation}
h_{\rho }=\sum_{k=1}^{n}\left( \ln (\rho _{k,k+1})+\frac{1}{2}\sum_{\lambda
=\pm 1}p_{k+(1+\lambda )/2}^{-1}\,\ln (\rho _{k,k+1})\,p_{k+(1+\lambda
)/2}+\gamma \right) .
\end{equation}
Here $\gamma =-\psi (1)$ is the Euler constant.

The invariance of the transfer matrix can be verified using two equivalent
representations for $q_{r}$: 
\begin{equation}
q_{r}=\sum_{i_{1}<i_{2}<...i_{r}}\prod_{l=1}^{r}\left(
\sum_{k=i_{l}+1}^{i_{l+1}}\rho _{k-1,k}\,\,\,\,p_{i_{l}}\right)
=\sum_{i_{1}<i_{2}<...i_{r}}\prod_{l=1}^{r}\left( \rho
_{i_{l},i_{l}+1}\,\sum_{k=i_{l}+1}^{i_{l+1}}p_{k}\right) \,.
\end{equation}

Note, that the supersymmetry corresponds to an analogous generalization of
translations to super-translations. Furthermore, the Kramers-Wannier duality
in the Ising model and the popular electromagnetic duality $\overrightarrow{E%
}\leftrightarrow \overrightarrow{H}$ can be considered as similar canonical
transformations [8].

The above duality symmetry is realized as an unitary transformation only for
the vanishing total momentum:

\begin{equation}
\overrightarrow{p}=\sum_{r=1}^{n}\overrightarrow{p_{r}}=0.
\end{equation}
In this case one can parametrize gluon momenta in terms of momentum
transfers $k_{r}$ as follows 
\begin{equation}
p_{r}=k_{r}-k_{r+1}\,,
\end{equation}
which gives a possibility to present the symmetry transformation in a
simpler form: 
\begin{equation}
k_{r}\rightarrow \rho _{r}\rightarrow k_{r+1}\,,\,r=1,2...n\,.
\end{equation}

Because the operators $q_{r}$ compose a complete set of invariants of the
transformation, the Hamiltonian $h$ should be their function 

\begin{equation}
h=h(q_{2},q_{3},...,q_{n})\,,
\end{equation}
fixed by the property of its locality. Furthermore, a common eigenfunction
of $q_{r}\,\,(r=2,...,n)$ is simultaneously a solution of the
Schr\"{o}dinger equation, which means, that the duality symmetry gives an
explanation of the integrability of the Reggeon model at $N_c \rightarrow
\infty$.

To formulate the duality as an integral equation we work in the
two-dimensional impact parameter space $\overrightarrow{\rho }$, initially
without taking into account the property of the holomorphic factorization of
the Green functions. The wave function $\psi _{m,\widetilde{m}}$ of the
composite state with $\overrightarrow{p}=0$ can be written in terms of the
eigenfunction $f_{m\widetilde{m}}$ of a commuting set of the operators $q_{k}
$ and $q_{k}^{*}$ for $k=1,2...n$ as follows

\begin{equation}
\psi _{m,\widetilde{m}}(\overrightarrow{\rho _{12}},\overrightarrow{\rho
_{23}},...,\overrightarrow{\rho _{n1}})=\int \frac{d^{2}\rho _{0}}{2\,\pi }%
\,f_{m,\widetilde{m}}(\overrightarrow{\rho _{1}},\overrightarrow{\rho _{2}}%
,...,\overrightarrow{\rho _{n}};\overrightarrow{\rho _{0}})\,.
\end{equation}

It is a highest-weight component of the M\"{o}bius group representation with
the quantum numbers $m=1/2+i\nu +n/2$ and $\,\widetilde{m}=1/2+i\nu -n/2$
related with eigenvalues of the Casimir operators

\begin{equation}
\left( \sum_{k=1}^{n}\overrightarrow{M_{k}}\right) ^{2}\psi _{m,\widetilde{m}%
}=m(m-1)\psi _{m,\widetilde{m}}\,\,\,,\,\,\,\,\,\,\left( \sum_{k=1}^{n}%
\overrightarrow{M_{k}^{*}}\right) ^{2}\psi _{m,\widetilde{m}}=\widetilde{m}(%
\widetilde{m}-1)\psi _{m,\widetilde{m}}\,.
\end{equation}

The other components of this highest-weight representation can be obtained
by applying to $\psi _{m,\widetilde{m}}$ the M\"{o}bius group generators:

\begin{equation}
\psi _{m,\widetilde{m}}^{r_1r_2}=\left( \sum_{k=1}^n\rho _k^2\,\partial
_k\right) ^{r_1}\left( \sum_{k=1}^n\rho _k^{*2}\,\partial _k^{*}\right)
^{r_2}\psi _{m,\widetilde{m}}\,.
\end{equation}

Note that, in accordance with the relations 
\begin{equation}
\widetilde{m}^{*}=1-m\,\,,\,\,\,m^{*}=1-\widetilde{m}
\end{equation}
the conjugate function $f_{m,\widetilde{m}}^{*}$ is transformed as $f_{1-m,1-%
\widetilde{m}}$:

\begin{equation}
\left( f_{m,\widetilde{m}}(\overrightarrow{\rho _{1}},...,\overrightarrow{%
\rho _{n}};\overrightarrow{\rho _{0}})\right) ^{*}\sim f_{1-m,1-\widetilde{m}%
}(\overrightarrow{\rho _{1}},...,\overrightarrow{\rho _{n}};\overrightarrow{%
\rho _{0}}).
\end{equation}
Moreover, because of the reality of the M\"{o}bius group, the
complex-conjugated representations are linearly dependent: 
\begin{equation}
\left( f_{m,\widetilde{m}}(\overrightarrow{\rho _{1}},...,\overrightarrow{%
\rho _{n}};\overrightarrow{\rho _{0}})\right) ^{*}\sim \int d^{2}\rho
_{0^{\prime }}\,(\rho _{00^{\prime }})^{2m-2}(\rho _{00^{\prime }}^{*})^{2%
\widetilde{m}-2}\,f_{m,\widetilde{m}}(\overrightarrow{\rho _{1}},...,%
\overrightarrow{\rho _{n}};\overrightarrow{\rho _{0^{\prime }}}).
\end{equation}

By considering the limit $\rho _{0}\rightarrow \infty $ of this equation, we
obtain for $\psi _{m,\widetilde{m}}$ the new representation 
\begin{equation}
\psi _{m,\widetilde{m}}(\overrightarrow{\rho _{12}},\overrightarrow{\rho
_{23}},...,\overrightarrow{\rho _{n1}})\sim f_{1-m,1-\widetilde{m}}(%
\overrightarrow{\rho _{1}},\overrightarrow{\rho _{2}},...,\overrightarrow{%
\rho _{n}};\infty )\,.
\end{equation}
Because of the relations 
\begin{equation}
\left( \psi _{\widetilde{m},m}(\overrightarrow{\rho _{12}},\overrightarrow{%
\rho _{23}},...,\overrightarrow{\rho _{n1}})\right) ^{*}\sim \psi _{1-%
\widetilde{m},1-m}(\overrightarrow{\rho _{12}},\overrightarrow{\rho _{23}}%
,...,\overrightarrow{\rho _{n1}})\,,\,
\end{equation}
the functions $\psi _{\widetilde{m},m}^{*}$ and $\psi _{m,\widetilde{m}}$
have the same conformal spin $n=m-\widetilde{m}$.

Taking into account the hermicity properties of the total Hamiltonian:

\begin{equation}
H^{+}=\prod_{k=1}^{n}\left| \rho _{k,k+1}\right| ^{-2}H\prod_{k=1}^{n}\left|
\rho _{k,k+1}\right| ^{2}=\prod_{k=1}^{n}\left| p_{k}\right|
^{2}H\prod_{k=1}^{n}\left| p_{k}\right| ^{-2},
\end{equation}
the solution $\psi _{\widetilde{m},m}^{+}$ of the complex-conjugated
Schr\"{o}dinger equation for $\overrightarrow{p}=0$ can be expressed in
terms of $\psi _{\widetilde{m},m}$ as follows :

\begin{equation}
\psi _{\widetilde{m},m}^{+}(\overrightarrow{\rho _{12}},\overrightarrow{\rho
_{23}},...,\overrightarrow{\rho _{n1}})=\prod_{k=1}^{n}\left| \rho
_{k,k+1}\right| ^{-2}\left( \psi _{\widetilde{m},m}(\overrightarrow{\rho
_{12}},\overrightarrow{\rho _{23}},...,\overrightarrow{\rho _{n1}})\right)
^{*}\,.
\end{equation}

If one performs the Fourier transformation of $\psi _{\widetilde{m},m}^{+}$
to the momentum space

\begin{equation}
\Psi _{m,\widetilde{m}}(\overrightarrow{p_{1}},\overrightarrow{p_{2}},...,%
\overrightarrow{p_{n}})=\int \prod_{k=1}^{n-1}\frac{d^{2}\rho
_{k-1,k}^{^{\prime }}}{2\pi }\prod_{k=1}^{n}e^{i\overrightarrow{p_{k}}\,%
\overrightarrow{\rho _{k}^{\prime }}}\psi _{\widetilde{m},m}^{+}(%
\overrightarrow{\rho _{12}^{\prime }},\overrightarrow{\rho _{23}^{\prime }}%
,...,\overrightarrow{\rho _{n1}^{\prime }})
\end{equation}
with substituting the arguments 

\begin{equation}
\overrightarrow{p_{k}}\rightarrow \overrightarrow{\rho _{k,k+1}}\,,
\end{equation}
the new expression $\Psi _{m,\widetilde{m}}(\overrightarrow{\rho _{12}},%
\overrightarrow{\rho _{23}},...,\overrightarrow{\rho _{n1}})$ will have the
same properties as the initial function $\psi _{m,\widetilde{m}}(%
\overrightarrow{\rho _{12}},\overrightarrow{\rho _{23}},...\overrightarrow{%
\rho _{n1}})$ under rotations and dilatations.

Moreover, in accordance with the above duality symmetry it satisfies the
same set of equations as $\psi _{m,\widetilde{m}}$ and therefore these two
functions can be chosen to be proportional:

\begin{equation}
\psi _{m,\widetilde{m}}(\overrightarrow{\rho _{12}},\overrightarrow{\rho
_{23}},...,\overrightarrow{\rho _{n1}})=c_{m,\widetilde{m}}\,\Psi _{m,%
\widetilde{m}}(\overrightarrow{\rho _{12}},\overrightarrow{\rho _{23}},...,%
\overrightarrow{\rho _{n1}})\,.
\end{equation}
The proportionality constant $c_{m,\widetilde{m}}$ is determined from the
condition that the norm of the function $\psi _{m,\widetilde{m}}$:

\begin{equation}
\left\| \psi _{m,\widetilde{m}}\right\| ^{2}=\int \prod_{k=1}^{n-1}\frac{%
d^{2}\rho _{k,k+1}}{2\,\pi }\,\,\psi _{m,\widetilde{m}}^{+}\psi _{m,%
\widetilde{m}}
\end{equation}
is conserved after this transformation.

Because $\psi _{m,\widetilde{m}}$ is also an eigenfunction of the integrals
of motion $A$ and $A^{*}$ with their eigenvalues $\lambda _{m}$ and $\lambda
_{m}^{*}=\lambda _{\widetilde{m}}$:

\begin{equation}
A\,\psi _{m,\widetilde{m}}=\lambda _{m}\,\psi _{m,\widetilde{m}%
}\,,\,\,\,\,A^{*}\,\psi _{m,\widetilde{m}}=\lambda _{\widetilde{m}}\,\psi
_{m,\widetilde{m}}\,,\,\,\,\,A=\rho _{12}...\rho _{n1}p_{1}...p_{n}\,,\,\,
\end{equation}
one can verify that, for the unitarity of the duality transformation, the
constant $c_{m,\widetilde{m}}$ should be chosen as

\begin{equation}
c_{m,\widetilde{m}}=\left| \lambda _{m}\right| \,\,2^{n}\,,
\end{equation}
for an appropriate phase of $\psi _{m,\widetilde{m}}$. Here the factor $2^{n}
$ appears due to the relation $\partial _{\mu }^{2}=4\partial \partial ^{*}$%
. This value of $c_{m,\widetilde{m}}$ is compatible also with the
requirement that two subsequent duality transformations are equivalent to
the cyclic permutation $i\rightarrow i+1$ of gluon indices.

Thus, the duality symmetry takes the form of the following integral equation
for $\psi _{m,\widetilde{m}}$: 
\begin{equation}
\psi _{m,\widetilde{m}}(\overrightarrow{\rho _{12}},...,\overrightarrow{\rho
_{n1}})=\left| \lambda _{m}\right| \,2^{n}\,\int \,\prod_{k=1}^{n-1}\frac{%
d^{2}\rho _{k-1,k}^{\prime }}{2\pi }\,\prod_{k=1}^{n}\frac{e^{i%
\overrightarrow{\rho _{k,k+1}}\,\overrightarrow{\rho _{k}^{\prime }}}}{%
\left| \rho _{k,k+1}^{\prime }\right| ^{2}}\,\psi _{\widetilde{m},m}^{*}(%
\overrightarrow{\rho _{12}^{\prime }},...,\overrightarrow{\rho _{n1}^{\prime
}})\,.
\end{equation}

Note that the validity of this equation in the Pomeron case $n=2$ can be
verified from the relations 
\[
f_{m,\widetilde{m}}(\overrightarrow{\rho _{1}},\overrightarrow{\rho _{2}};%
\overrightarrow{\rho _{0}})\sim \left( \frac{\rho _{12}}{\rho _{10}\rho _{20}%
}\right) ^{m}\left( \frac{\rho _{12}^{*}}{\rho _{10}^{*}\rho _{20}^{*}}%
\right) ^{\widetilde{m}}\,\,,\,\,\,\,\,\,\psi _{m,\widetilde{m}}(%
\overrightarrow{\rho _{12}})\sim (\rho _{12})^{1-m}(\rho _{12}^{*})^{1-%
\widetilde{m}}\,\,, 
\]
\[
\left| \lambda _{m}\right| \,2^{2}\,\int \,\frac{d^{2}\rho _{12}^{\prime }}{%
2\pi }\,\frac{e^{i\overrightarrow{\rho _{12}}\,\overrightarrow{\rho
_{12}^{\prime }}}}{\left| \rho _{12}^{\prime }\right| ^{4}}\,(\rho
_{12}^{\prime })^{\widetilde{m}}(\rho _{12}^{\prime *})^{m}=e^{i\delta (m,%
\widetilde{m})}{\rho _{12}}^{1-m}\,{\rho _{12}^{*}}^{1-\widetilde{m}%
},\,\,\,\left| \lambda _{m}\right| =\left| m(1-m)\right| , 
\]
\begin{equation}
e^{i\delta (m,\widetilde{m})}=2^{m+\widetilde{m}-1}(-i)^{\widetilde{m}-m}%
\frac{\Gamma (1+m)}{\Gamma (2-\widetilde{m})}\,\left( \frac{m^{*}(m^{*}-1)}{%
m(m-1)}\right) ^{1/2}\,.
\end{equation}

Let us use for $f_{m,\widetilde{m}}$ the conformally covariant anzatz 
\begin{equation}
f_{m,\widetilde{m}}(\overrightarrow{\rho _{1}},...,\overrightarrow{\rho _{n}}%
;\overrightarrow{\rho _{0}})=\left( \prod_{i=1}^{n}\frac{\rho _{i,i+1}}{\rho
_{i0}^{2}}\right) ^{m/n}\left( \prod_{i=1}^{n}\frac{\rho _{i,i+1}^{*}}{\rho
_{i0}^{*2}}\right) ^{\widetilde{m}/n}f_{m,\widetilde{m}}(\overrightarrow{%
x_{1}},...,\overrightarrow{x_{n}})\,,
\end{equation}
where the anharmonic ratios $x_{r}$ ($r=1,2...n$) of the gluon coordinates
are chosen as follows

\begin{equation}
x_{r}=\frac{\rho _{r-1,r}\,\rho _{r+1,0}}{\rho _{r-1,0}\,\rho _{r+1,r}}%
\,;\,\,\,\,\,\,\,\prod_{r=1}^{n}x_{r}=(-1)^{n}\,;\,\,\,\,\,%
\sum_{r=1}^{n}(-1)^{r}\prod_{k=r+1}^{n}x_{k}=0\,.
\end{equation}
The function $f_{m,\widetilde{m}}(\overrightarrow{x_{1}},...,\overrightarrow{%
x_{n}})$ is invariant under certain modular transformations as a consequence
of the Bose symmetry.

For the physically interesting case $m,\widetilde{m}\rightarrow 1/2$ we can
calculate $\psi _{m,\widetilde{m}}$, taking into account the logarithmic
divergence of the integral at $\overrightarrow{\rho _{0}}\rightarrow \infty $%
:

\begin{equation}
\lim_{m,\widetilde{m}\rightarrow 1/2}\psi _{m,\widetilde{m}}(\overrightarrow{%
\rho _{12}},\overrightarrow{\rho _{23}},...\overrightarrow{\rho _{n1}})=%
\frac{f(\overrightarrow{z_1},...\overrightarrow{z_n})}{1-m-\widehat{m}}%
\,\prod_{k=1}^n\left| \rho _{k,k+1}\right|^{1/n} \,,\,\,z_r= \frac{\rho
_{r-1,r}\,}{\,\rho _{r+1,r}}\,.
\end{equation}
Let us change the variables $z_r$ to new ones $y_{k}$ as follows:

\begin{equation}
y_{k}=\rho _{k+1,k}/\rho
_{n,n-1}=(-1)^{n-k-1}\prod_{r=k+1}^{n-1}z_{r}\,\,,\,\,\,\,\,y_{n-1}=1\,,\,\,%
\,\sum_{k=1}^{n}y_{k}=0\,.
\end{equation}

The duality equation for the wave function $f(\overrightarrow{y_{1}},...,%
\overrightarrow{y_{n-2}})$ can be written in the form:

\begin{equation}
f(\overrightarrow{y_{1}},...,\overrightarrow{y_{n-2}})=\left| \lambda
\right| \,\,\int \prod_{k=1}^{n-2}\frac{d^{2}y_{k}^{\prime }}{2\pi }\,K(%
\overrightarrow{y};\overrightarrow{y^{\prime }})\prod_{k=1}^{n}\frac{\left|
y_{k}\right| ^{1/n}}{\left| y_{k}^{\prime }\right| ^{2-1/n}}\,\,f(%
\overrightarrow{y_{1}^{\prime }},...,\overrightarrow{y_{n-2}^{\prime }})\,.
\end{equation}
The integral kernel $K(\overrightarrow{y};\overrightarrow{y^{\prime }})$ is
given below:

\begin{equation}
K(\overrightarrow{y};\overrightarrow{y^{\prime }})=2^{n}\int \frac{d^{2}\rho
_{n,n-1}^{\prime }}{2\pi \left| \rho _{n,n-1}^{\prime }\right| ^{3}\left|
\rho _{n,n-1}\right| }\,\exp \left( -i\,\sum_{k=1}^{n}\left( \overrightarrow{%
\rho _{k}}\,,\,\overrightarrow{y_{k-1}^{\prime }\rho _{n,n-1}^{^{\prime }}}%
\right) \right) \,
\end{equation}
and is calculated analytically:

\begin{equation}
K(\overrightarrow{y};\overrightarrow{y^{\prime }})=-2^{n}\left|
\sum_{k=1}^{n}y_{k-1}^{\prime }\rho _{k}^{*}\right| \left| \rho
_{n,n-1}\right| ^{-1}=-2^{n}\left| \sum_{k=1}^{n-1}y_{k-1}^{\prime
}\sum_{r=n}^{k-1}y_{r}^{*}\right| .
\end{equation}

To simplify the duality equation, we consider below the compound state of
three reggeized gluons.

\section{Duality equation for the Odderon wave function}

In the case of the Odderon the conformal invariance fixes the solution of
the Schr\"{o}dinger equation [5]

\begin{equation}
f_{m,\widetilde{m}}(\overrightarrow{\rho _1},\overrightarrow{\rho _2},%
\overrightarrow{\rho _3};\overrightarrow{\rho _0})=\left( \frac{\rho
_{12}\,\rho _{23}\,\rho _{31}}{\rho _{10}^2\,\rho _{20}^2\,\rho _{30}^2}%
\right) ^{m/3}\left( \frac{\rho _{12}^{*}\,\rho _{23}^{*}\,\rho _{31}^{*}}{%
\rho _{10}^{*2}\,\rho _{20}^{*2}\,\rho _{30}^{*2}}\right) ^{\widetilde{m}%
/3}f_{m,\widetilde{m}}(\overrightarrow{x})\,
\end{equation}
up to an arbitrary function $f_{m,\widetilde{m}}(\overrightarrow{x})$ of one
complex variable $x$ being the anharmonic ratio of four coordinates\thinspace

\begin{equation}
x=\frac{\rho _{12}\,\rho _{30}}{\rho _{10}\,\rho _{32}}\,.
\end{equation}

Note that, owing to the Bose symmetry of the Odderon wave function, $f_{m,%
\widetilde{m}}(\overrightarrow{x})$ has the following modular properties:

\begin{equation}
f_{m,\widetilde{m}}(\overrightarrow{x})=(-1)^{(\widetilde{m}-m)/3}\,f_{m,%
\widetilde{m}}(\overrightarrow{x}/\left| x\right| ^{2})\,=(-1)^{(\widetilde{m%
}-m)/3}\,f_{m,\widetilde{m}}(\overrightarrow{1}-\overrightarrow{x})\,
\end{equation}
and satisfies the normalization condition 
\begin{equation}
\left\| f_{m,\widetilde{m}}\right\| ^{2}= \int \frac{d^{2}x}{\left|
x(1-x)\right| ^{4/3}}\left| f_{m,\widetilde{m}}(\overrightarrow{x})\right|
^{2}\,,
\end{equation}
compatible with the modular symmetry.

After changing the integration variable from $\rho _{0}$ to $x$ in
accordance with the relations

\begin{equation}
\rho _{0}=\frac{\rho _{31}\rho _{12}}{\rho _{12}-x\,\rho _{32}}+\rho
_{1}\,,\,\,\,\,\,\,\,d\,\rho _{0}=\frac{\rho _{31}\rho _{12}\rho _{32}}{%
(\rho _{12}-x\,\rho _{32})^{2}}\,\,dx
\end{equation}
the Odderon wave function $\psi _{m,\widetilde{m}}(\overrightarrow{\rho _{ij}%
})$ at $\overrightarrow{q}=0$ can be written as

\begin{equation}
\psi _{m,\widetilde{m}}(\overrightarrow{\rho _{ij}})=\left( \frac{\rho _{23}%
}{\rho _{12}\rho _{31}}\right) ^{m-1}\left( \frac{\rho _{23}^{*}}{\rho
_{12}^{*}\rho _{31}^{*}}\right) ^{\widetilde{m}-1} \chi _{m,\widetilde{m}}(%
\overrightarrow{z})\,\,,\,\,\,z=\frac{\rho _{12}}{\rho _{32}}\,,
\end{equation}
where

\begin{equation}
\chi _{m,\widetilde{m}}(\overrightarrow{z})=\int \frac{d^{2}x\,\,f_{m,%
\widetilde{m}}(\overrightarrow{x})}{2\,\pi \left| x-z\right| ^{4}}\,\left( 
\frac{(x-z)^{3}}{x(1-x)}\right) ^{2m/3}\left( \frac{(x^{*}-z^{*})^{3}}{%
x^{*}(1-x^{*})}\right) ^{2\widetilde{m}/3}.
\end{equation}
In fact this function is proportional to $f_{1-m,1-\widetilde{m}}(%
\overrightarrow{z})$: 
\begin{equation}
\chi _{m,\widetilde{m}}(\overrightarrow{z})\sim \left( x(1-x)\right)
^{2(m-1)/3}\left( x^{*}(1-x^{*})\right) ^{2(\widetilde{m}-1)/3}f_{1-m,1-%
\widetilde{m}}(\overrightarrow{z})\,\,,
\end{equation}
which is a realization of the discussed linear dependence between two
representations $(m,\widetilde{m})$ and $(1-m,1-\widetilde{m})$.

The corresponding reality property for the M\"{o}bius group representations
can be presented in the form of the integral relation 
\begin{equation}
\chi _{m,\widetilde{m}}(\overrightarrow{z})=\int \frac{d^{2}x}{2\,\pi }%
\,(x-z)^{2m-2}\,(x^{*}-z^{*})^{2\widetilde{m}-2}\,\chi _{1-m,1-\widetilde{m}%
}(\overrightarrow{x})\,
\end{equation}
for an appropriate choice of phases of the functions $\chi _{m,\widetilde{m}%
} $ and $\chi _{1-m,1-\widetilde{m}}$. The function $\chi _{m,\widetilde{m}}(%
\overrightarrow{z})$ satisfies the modular relations

\begin{equation}
\chi _{m,\widetilde{m}}(\overrightarrow{z})=(-1)^{m-\widetilde{m}%
}z^{2m-2}z^{*2\widetilde{m}-2}\chi _{m,\widetilde{m}}(\overrightarrow{z}%
/\left| z\right| ^2)=(-1)^{m-\widetilde{m}}\chi _{m,\widetilde{m}}(%
\overrightarrow{1}-\overrightarrow{z})\,.
\end{equation}

The duality property of the wave function $\psi _{m,\widetilde{m}}(%
\overrightarrow{\rho _{ij}})$ can be written in a form of the integral
equation: 
\begin{equation}
\psi _{m,\widetilde{m}}(\overrightarrow{\rho _{ij}})=\left| \lambda
_{m}\right| \,2^{3}\,\int \frac{d^{2}\rho _{12}^{\prime }}{2\,\pi }\frac{%
d^{2}\rho _{23}^{\prime }}{2\,\pi }\frac{\exp (i(\overrightarrow{\rho
_{21}^{\prime }}\,\overrightarrow{\rho _{23}}+\overrightarrow{\rho
_{31}^{\prime }}\,\overrightarrow{\rho _{31}}))}{\left| \rho _{12}^{\prime
}\rho _{23}^{\prime }\rho _{31}^{\prime }\right| ^{2}}\,\psi _{\widetilde{m}%
,m}^{*}(\overrightarrow{\rho _{ij}^{\prime }})\,,
\end{equation}
where $\left| \lambda _{m}\right| ^{2}$ is the corresponding eigenvalue of
the differential operator

\begin{equation}
\left| A\right| ^2=\left| \rho _{12}\rho _{23}\rho _{31}\,p_1p_2p_3\right|
^2\,.
\end{equation}
The function $\psi _{\widetilde{m},m}^{*}(\overrightarrow{\rho _{ij}^{\prime
}})$ is transformed as $\psi _{1-\widetilde{m},1-m}(\overrightarrow{\rho
_{ij}^{\prime }})$:

\begin{equation}
\psi _{\widetilde{m},m}^{*}(\overrightarrow{\rho _{ij}^{\prime }})\sim \psi
_{1-\widetilde{m},1-m}(\overrightarrow{\rho _{ij}^{\prime }})\,.
\end{equation}

In terms of $\chi _{m,\widetilde{m}}(\overrightarrow{z})$ the above duality
equation looks as follows

\begin{equation}
\frac{\chi _{m,\widetilde{m}}(\overrightarrow{z})}{\left| \lambda
_{m}\right| }=\int \frac{d^{2}z^{\prime }\left( z^{\prime }(1-z^{\prime
})\right) ^{\widetilde{m}-1}\left( z^{\prime *}(1-z^{\prime *})\right) ^{m-1}%
}{2\pi \left( z(1-z)\right) ^{1-m}\left( z^{*}(1-z^{*})\right) ^{1-%
\widetilde{m}}}\,R(\overrightarrow{z},\overrightarrow{z^{\prime }})\chi _{%
\widetilde{m},m}^{*}(\overrightarrow{z^{\prime }})\,,
\end{equation}
where

\begin{equation}
R(\overrightarrow{z},\overrightarrow{z^{\prime }})=\int \frac{8\,d^{2}\rho
_{32}^{\prime }/(2\pi )}{\left| \rho _{32}^{\prime }\right| ^{4}\left| \rho
_{32}\right| ^{2}}\left( \rho _{32}\rho _{32}^{\prime *}\right) ^{m}\left(
\rho _{32}^{*}\rho _{32}^{\prime }\right) ^{\widetilde{m}}\exp (i(%
\overrightarrow{\rho _{32}}\overrightarrow{\rho _{12}^{\prime }}+%
\overrightarrow{\rho _{13}}\overrightarrow{\rho _{13}^{\prime }}))\,.
\end{equation}

This integral is calculated to be 
\begin{equation}
R(\overrightarrow{z},\overrightarrow{z^{\prime }})=C_{m,\widetilde{m}%
}\,Z^{1-m}\,(Z^{*})^{1-\widetilde{m}}
\end{equation}
with 
\begin{equation}
C_{m,\widetilde{m}}=\frac{2\,e^{i\delta (m,\widetilde{m})}}{\left|
m(1-m)\right| }=-2^{\widetilde{m}+m}\frac{(-i)^{\widetilde{m}-m}}{\sin (\pi
m)}\frac{\Gamma (1/2)}{\Gamma (2-m)}\frac{\Gamma (1/2)}{\Gamma (2-\widetilde{%
m})}\,,
\end{equation}
where the phase $\delta (m,\widetilde{m})$ was defined above and

\begin{equation}
Z=zz^{\prime *}-z+1=z(z^{\prime *}-1+\frac{1}{z})\,.
\end{equation}

Further, we introduce the new integration variable $z^{\prime }\rightarrow
z^{\prime }{}^{*}$, which effectively leads to the substitution $\chi _{%
\widetilde{m}.m}(\overrightarrow{z^{\prime }})\rightarrow \chi _{m,%
\widetilde{m}}(\overrightarrow{z^{\prime }})$. By performing the modular
transformation $1-\frac{1}{z}\rightarrow z$ and taking into account the
modular relation 
\begin{equation}
\chi _{m,\widetilde{m}}(\overrightarrow{z})=(1-z)^{2m-2}(1-z^{*})^{2%
\widetilde{m}-2}\chi _{m,\widetilde{m}}((\overrightarrow{1}-\overrightarrow{z%
})/\left| 1-z\right| ^{2})\,,
\end{equation}
one can rewrite the above duality equation for $\chi _{m,\widetilde{m}}(%
\overrightarrow{z})$ as 
\begin{equation}
\frac{\chi _{m,\widetilde{m}}(\overrightarrow{z})}{\left| \lambda
_{m}\right| \,C_{m,\widetilde{m}}}=\int \frac{d^{2}z^{\prime }}{2\pi }\left( 
\frac{z^{\prime }(1-z^{\prime })z(1-z)}{z^{\prime }-z}\right) ^{m-1}\left( 
\frac{z^{\prime }(1-z^{\prime })z(1-z)}{z^{\prime }-z}\right) ^{*\,%
\widetilde{m}-1}\chi _{m,\widetilde{m}}^{*}(\overrightarrow{z^{\prime }}).
\end{equation}
This equation for $\chi _{m,\widetilde{m}}(\overrightarrow{z})$ corresponds
to the symmetry of the Odderon wave function under the involution $%
p_{k}\leftrightarrow \frac{1}{2}\varepsilon _{klm}\rho _{lm}$.

It can be written in the pseudo-differential form: 
\begin{equation}
z(1-z)\,(i\partial )^{2-m}\,z^{*}(1-z^{*})\,(i\partial ^{*})^{2-\widetilde{m}%
}\,\varphi _{1-m,1-\widetilde{m}}(\overrightarrow{z})=\left| \lambda _{m,%
\widetilde{m}}\right| \,\left( \varphi _{1-m,1-\widetilde{m}}(%
\overrightarrow{z})\right) ^{*}\,,
\end{equation}
where 
\begin{equation}
\varphi _{1-m,1-\widetilde{m}}(\overrightarrow{z})=2^{(1-m-\widetilde{m}%
)/2}\,\left( z(1-z)\right) ^{1-m}\left( z^{*}(1-z^{*})\right) ^{1-\widetilde{%
m}}\,\chi _{m,\widetilde{m}}(\overrightarrow{z})\,.
\end{equation}
Note, that for a self-consistency of the above equation its right-hand side
should be ortogonal to the zero modes of the operator $(i\partial
)^{2-m}(i\partial ^{*})^{2-\widetilde{m}}$. Due to the Bose symmetry of the
wave function it is enough to impose on it only one integral constraint: 
\begin{equation}
\int \frac{d^{2}z}{|z(1-z)|^{2}}\,\left( \varphi _{1-m,1-\widetilde{m}}(%
\overrightarrow{z})\right) ^{*}=0\,.
\end{equation}

The normalization condition for the wave function

\[
\left\| \varphi _{m,\widetilde{m}}\right\| ^{2}=\int \frac{d^{2}x}{\left|
x(1-x)\right| ^{2}}\,\left| \varphi _{m,\widetilde{m}}(\overrightarrow{x}%
)\right| ^{2} 
\]
is compatible with the duality symmetry.

The holomorphic and anti-holomorphic factors of $f_{m,\widetilde{m}}(%
\overrightarrow{\rho _{1}},\overrightarrow{\rho _{2}},\overrightarrow{\rho
_{3}};\overrightarrow{\rho _{0}})$ are eigenfunctions of the integrals of
motion $A$ and $A^{*}$:

\begin{equation}
A\,f_{m}=\lambda _{m}\,f_{m}\,,\,\,\,\,A^{*}\,f_{\widetilde{m}}=\lambda _{%
\widetilde{m}}\,f_{\widetilde{m}}\,\,,\,\,\,\,\lambda _{\widetilde{m}%
}=(\lambda _{m})^{*}\,.
\end{equation}

In the limit $m\rightarrow 1/2,\,\widetilde{m}\rightarrow 1/2$,
corresponding to the ground state, the integral over $x$ for the wave
function $\psi _{m,\widetilde{m}}$ at $q=0$ is calculated, since the main
contribution appears from the singularity at $x=z$:

\begin{equation}
\psi _{m,\widetilde{m}}(\overrightarrow{\rho _{ij}})\simeq \frac{1}{(m+%
\widetilde{m}-1)}\,\left| \rho _{12}\rho _{23}\rho _{31}\right|
^{1/3}f(z,z^{*})\,\,,\,\,\,\,z=\rho _{12}/\rho _{32}\,,
\end{equation}
where $f(x,x^{*})=f_{1/2,1/2}(x,x^{*})$. This means that one can obtain for $%
f(x,x^{*})$ the following equation in the $x$-representation:

\begin{equation}
\left| x(1-x)\right| ^{1/3}f(x,x^{*})=-\frac{4\left| \lambda \right| }{\pi }%
\int \frac{d^{2}y\,\,\left| y-x\right| }{\left| y(1-y)\right| ^{5/3}}%
\,\,f(y,y^{*})\,
\end{equation}
for a definite choice of the phase of the ground-state function $f$.

The common factor in front of the integral is in agreement with the
condition of conservation of the norm of the wave function after this
canonical transformation (only its sign can be changed). The above integral
relation is reduced to the following pseudo-differential equation

\begin{equation}
\left| x(1-x)\,p^{3/2}\right| ^{2}\varphi (x,x^{*})=\left| \lambda \right|
\,\,\,\varphi (x,x^{*})\,,
\end{equation}
where $p=i\,\partial /\partial (x)\,,\,p^{*}=i\,\partial /\partial (x^{*})$,
if we introduce the new function

\begin{equation}
\varphi (x,x^{*})=\left| x(1-x)\right| ^{1/3}f(x,x^{*})
\end{equation}
and use the relation 
\begin{equation}
\left| p\right| ^{3}\,\left| x\right| =-\left| p\right| ^{3}\int_{-\pi
/2}^{\pi /2}\frac{d\phi }{2\pi }\,\,\int_{-\infty }^{\infty }\frac{d\left| 
\overrightarrow{p}\right| }{\left| \overrightarrow{p}\right| ^{2}}\,\,\exp
(-i\overrightarrow{p}\overrightarrow{x})=-\frac{\pi }{4}\,\,\delta ^{2}(%
\overrightarrow{x})\,.
\end{equation}
For self-consistency of the pseudo-differential equation one should impose
on the wave function the following constraint: 
\begin{equation}
\int \,\frac{d^{2}x}{|x(1-x)|^{2}}\,\varphi (x,x^{*})=0\,.
\end{equation}

Let us derive the duality equation for $\varphi _{m,\widetilde{m}}(%
\overrightarrow{x})$ for general values of $m$ and $\widetilde{m}$ by using
different arguments. We start with the conformally covariant anzatz for the
holomorphic factor $\,f_{m}(\rho _{1},\rho _{2},\rho _{3};\rho _{0})$:

\begin{equation}
f_{m}(\rho _{1},\rho _{2},\rho _{3};\rho _{0})=\left( \frac{\rho _{12}\,\rho
_{23}\,\rho _{31}}{\rho _{10}^{2}\,\,\rho _{20}^{2}\,\,\rho _{30}^{2}}%
\right) ^{m/3}f_{m}(x)\,\,,\,\,\,\,x=\frac{\rho _{12}\,\rho _{30}}{\rho
_{10}\,\,\rho _{32}}\,.
\end{equation}

In the $x$ representation the integral of motion $A=\rho _{12}\,\rho
_{23}\,\rho _{31}\,p_{1}p_{2}p_{3}$ is an ordinary differential operator.
When acting on $f_m (x)$, it can be presented in the following form 
\[
\frac{i^{3}}{x}\left( \frac{m}{3}(x-2)+x(1-x)\partial \right) \frac{x}{1-x}%
\left( \frac{m}{3}(1+x)+x(1-x)\partial \right) \frac{1}{x}\left( \frac{m}{3}%
(1-2x)+x(1-x)\partial \right) 
\]
\begin{equation}
=X^{-m/3}\,A_{m}\,\,X^{m/3}\,\,,\,\,\,
X=x\,(1-x)\,\,,\,\,\,\,A_{m}=a_{1-m}\,a_{m}\,\,,
\end{equation}
where 
\begin{equation}
a_{m}=\,x\,\,(1-x)\,p^{m+1},\,\,\,\,\,\,a_{1-m}=\,x\,\,(1-x)\,\,p^{2-m}\,,\,%
\,\,\,\,\,p=i\,\partial \,.
\end{equation}

Therefore the differential equation $A\,\,f_{m}=\lambda _{m}\,\,f_{m}$ for
eigenfunctions $f_{m}$ and eigenvalues $\lambda _{m}$ is equivalent to the
system of two pseudo-differential equations 
\begin{equation}
a_{m}\,\varphi _{m}=l_{m}\,\,\varphi
_{1-m}\,\,,\,\,\,\,\,\,\,a_{1-m}\,\varphi _{1-m}=l_{1-m}\,\varphi _{m}\,,
\end{equation}
where we introduced the functions $\varphi _{m}$ and $\varphi _{1-m}$ in
accordance with the definitions 
\begin{equation}
\varphi _{m}\equiv X^{m/3}\,f_{m}\,,\,\,\,\,\,\,\varphi _{1-m}\equiv
X^{(1-m)/3}\,f_{1-m}\,
\end{equation}
and the eigenvalues $l_{m}$ and $l_{1-m}$ are related with the eigenvalue $%
\lambda _m$ of $A_m$ as follows 
\begin{equation}
\,\,l_{1-m}\,l_{m}=\lambda _{m}.
\end{equation}

The function $\varphi _{1-m}(x)$ has the conformal weight equal to $1-m$. It
is important that there is another relation between $\varphi _{m}$ and $%
\varphi _{1-m}$%
\begin{equation}
\varphi _{1-m}=R_{m}(x,p)\,\varphi _{m}\,\,,
\end{equation}
where 
\begin{equation}
R_{m}(x,p)=X^{-m+1}\,p^{1-2m}\,X^{-m}\,,\,\,\,\,\,\,R_{1-m}(x,p)=\left(
R_{m}(x,p)\right) ^{-1}\,.
\end{equation}
This follows from the fact that, for the M\"{o}bius group, the
complex-conjugated representations $O_{m,\widetilde{m}}(\overrightarrow{\rho
_{0}})$ and $O_{1-m,1-\widetilde{m}}(\overrightarrow{\rho _{0}})$ are
linearly dependent. To obtain the above expression for $R_m(x,p)$ one should
use for the Odderon wave function $f_{m,\widetilde{m}}(\overrightarrow{\rho
_{1}},\overrightarrow{\rho _{2}},\overrightarrow{\rho _{3}};\overrightarrow{%
\rho _{0}})$ the coformal anzatz and the property of holomorphic
factorization. The transformation

\begin{equation}
\varphi ^{\prime }(\overrightarrow{x})=U\,\varphi (\overrightarrow{x}%
)\,,\,\,\,\,\,\,U=R_{m}(x,p)\,R_{\widetilde{m}}(x^{*},p^{*})
\end{equation}
is unitary for our choice of the norm:

\begin{equation}
\left\| \varphi ^{\prime }\right\| ^{2}=\left\| \varphi \right\| ^{2}\,.
\end{equation}

Because

\[
\int \frac{d^{2}x}{\left| x(1-x)\right| ^{2}}\,(\varphi (\overrightarrow{x}%
))^{*}\,\,A_{m}\,(x)\,A_{\widetilde{m}}(x^{*})\,\varphi (\overrightarrow{x}%
)=\left\| A_{m}\,\varphi \right\| ^{2}\,, 
\]
the eigenvalue of the operator $A_{\widetilde{m}}(x^{*})$ $\,$is complex
conjugated to $\lambda _{m}$: 
\begin{equation}
A_{m}(x)\,\,\varphi (\overrightarrow{x})=\lambda _{m}\,\,\varphi (%
\overrightarrow{x})\,\,,\,\,\,\,\,\,A_{\widetilde{m}}(x^{*})\,\varphi (%
\overrightarrow{x})=\lambda _{m}^{*}\,\varphi (\overrightarrow{x})\,.
\end{equation}

Note that due to its M\"{o}bius covariance $\varphi (\overrightarrow{x})$ is
the sum of products of the eigenfunctions having opposite signs of their
eigenvalues $\lambda $:

\[
\varphi (\overrightarrow{x})=\sum C_{ik}\left( \varphi _{m,\lambda
}^{i}(x)\,\varphi _{\widetilde{m},\lambda ^{*}}^{k}(x^{*})+\varphi
_{m,-\lambda }^{i}(x)\,\varphi _{\widetilde{m},-\lambda
^{*}}^{k}(x^{*})\,\right) \,\,. 
\]
Since under the simultaneous transformations $x\leftrightarrow x^{*}$ and$%
\,\,m\leftrightarrow \widetilde{m}$ this function should be symmetric for
fixed $\lambda $, we conclude, that eigenvalues satisfy one of two relations

\begin{equation}
\lambda _{m}=\pm (\lambda _{m})^{*}
\end{equation}
and therefore $\lambda _{m}$ can be purely real or imaginary. It turns out,
that $\lambda _{m}$ is imaginary as a consequence of the modular invariance
[9].

One can veryfy the validity of the following relation

\begin{equation}
R_{m}(x,p)\,a_{1-m}\,a_{m}=a_{m}\,a_{1-m}\,R_{m}(x,p)\,,
\end{equation}
if the following identity is used:

\[
X^{-m}a_{1-m}a_{m}X^{m}=x^{2}(1-x)^{2}p^{3}+2(1+m)\,x(1-x)\left(
i\,p\,(1-2x)\,p-p\right) + 
\]
\begin{equation}
m(1+m)\,\left( x(1-x)\,p-(1-2x)^{2}\,p+i\,m\,(1-2x)\right) .
\end{equation}

In particular this relation means that two eigenvalues of $A_{m}$ coincide

\begin{equation}
\lambda _{m}=\lambda _{1-m}\,\,
\end{equation}
for the eigenfunctions $\varphi _{m}$ and $\varphi _{1-m}$ linearly related
by $R_{m}(x,p)$. Let us introduce the operators 
\begin{equation}
S_{m}(x,p)=R_{1-m}(x,p)\,a_{m}\,,\,\,\,\,\,\,S_{1-m}^{t}(x,p)=a_{1-m}%
\,R_{m}(x,p)\,,\,
\end{equation}
leaving the function $\varphi _{m}$ in the same space. Due to the above
formulas these operators commute with one another: 
\begin{equation}
A_{m}=S_{1-m}^{t}(x,p)\,S_{m}(x,p)=S_{m}(x,p)\,S_{1-m}^{t}(x,p)\,.
\end{equation}

Therefore we obtain two duality equations for the wave function whose
conformal weight is equal to $m$: 
\begin{equation}
S_{m}(x,p)\,\varphi _{m}=\,L_{1}^{(m)}\,\varphi
_{m}\,,\,\,\,\,\,\,S_{1-m}^{t}(x,p)\,\varphi _{m}=L_{2}^{(m)}\,\varphi
_{m}\,.
\end{equation}
As a consequence of the relation 
\begin{equation}
L_{1}^{(m)}L_{2}^{(m)}=\lambda \,
\end{equation}
between the eigenvalues of $S_m$ and $S_{1-m}^t$, the second equation
follows from the first one. 
%

In the particular case $m=1/2$, the equation for the holomorphic factors $%
\varphi (x)$ looks especially simple:

\begin{equation}
x(1-x)p^{3/2}\varphi _{\pm \sqrt{\lambda }}(x)=\pm \sqrt{\lambda }\varphi
_{\pm \sqrt{\lambda }}(x)
\end{equation}
and can be reduced in the $p$-representation to the Schr\"odinger equation
with the potential $V(p)=p^{-3/2}$. For $m=\widetilde{m}=1/2$ the total
odderon wave function $\varphi (x,x^{*})$, symmetric under the above
canonical transformation, is a solution of the equation:

\begin{equation}
\left| x(1-x)\right|^{2}\,|p|^3 \,\varphi (x,x^{*})=\left| \lambda \right|
\,\varphi (x,x^{*}),
\end{equation}
where in accordance with its hermicity properties the eigenvalue of the
operator $\sqrt{A^{*}}$ for the anti-holomorphic factor $\varphi (x^{*})$ is
taken to be equal to $\lambda ^{*}$.

\section{Single-valuedness condition}

There are three independent solutions $\varphi _{i}^{(m)}(x,\lambda )$ of
the third-order ordinary differential equation

\begin{equation}
a_{1-m}\,a_{m}\,\varphi =-ix(1-x)\left( x(1-x)\partial ^{2}+(2-m)\left(
(1-2x)\partial -1+m\right) \right) \partial \,\varphi =\lambda \,\,\varphi
\end{equation}
for each eigenvalue $\lambda $. In the region $x\rightarrow 0$ they can be
chosen as follows

\begin{equation}
\varphi _{r}^{(m)}(x,\lambda )=\sum_{k=1}^{\infty }d_{k}^{(m)}(\lambda
)\,\,x^{k}\,\,,\,\,\,\,\,d_{1}^{(m)}(\lambda )=1\,.
\end{equation}
\begin{equation}
\varphi _{s}^{(m)}(x,\lambda )=\sum_{k=0}^{\infty }a_{k}^{(m)}(\lambda
)\,\,x^{k}+\,\varphi _{r}^{(m)}(x,\lambda )\,\ln
x\,,\,\,\,\,\,a_{1}^{(m)}=0\,,
\end{equation}
\begin{equation}
\varphi _{f}^{(m)}(x,\lambda )=\sum_{k=0}^{\infty }c_{k+m}^{(m)}(\lambda
)\,\,x^{k+m}\,,\,\,\,\,\,c_{m}^{(m)}(\lambda )=1\,.
\end{equation}
The appearance of $\ln x$ in $\varphi _{s}^{(m)}(x,\lambda )$ is related
with the degeneracy of the differential equation in the small-$x$ region.
There is an ambiguity in the definition of $\varphi _{s}^{(m)}(x,\lambda )$
because one can add to it the function $\varphi _{r}^{(m)}(x,\lambda )$ with
an arbitrary coefficient. We have chosen $a_{1}^{m}=0$ to remove this
uncertainty.

Taking into account that for $\rho _{1}\rightarrow \rho _{2}$ the operator
product expansion is applicable, the functions $\varphi _{i}^{(m)}(x,\lambda
)$ can be considered as contributions of the holomorphic composite operators 
$O_{i}^{(M)}(\rho _{1})$ with the conformal weights $M=0,m$ and $1$ for $%
i=s,\,f$ and $r$ correspondingly. In this interpretation the above
degeneracy is related with the existence of the conserved vector current
(for $m=1/2$ there is also a conserved fermion current).

Due to the above differential equation the coefficients $a,c$ and $d$
satisfy the following recurrence relations 
\begin{eqnarray*}
i\lambda a_{k}^{(m)} &=&\left( a_{k+1}^{(m)}+d_{k+1}^{(m)}\frac{d}{d\,k}%
\right) k(k+1)(k+1-m)-\left( a_{k}^{(m)}+d_{k}^{(m)}\frac{d}{d\,k}\right)
k(k-m)(2k-m) \\
&&+\left( a_{k-1}^{(m)}+d_{k-1}^{(m)}\frac{d}{d\,k}\right)
(k-1)(k-m)(k-1-m)\,\,,
\end{eqnarray*}
\begin{eqnarray*}
i\lambda c_{k+m}^{(m)}
&=&(k+m)(k+m+1)(k+1)c_{k+m+1}^{(m)}-(k+m)k(2k+m)c_{k+m}^{(m)} \\
&&+(k+m-1)k(k-1)c_{k+m-1}^{(m)}\,,
\end{eqnarray*}
\begin{eqnarray}
i\lambda d_{k}^{(m)} &=&k(k+1)(k+1-m)d_{k+1}^{(m)}-k(k-m)(2k-m)d_{k}^{(m)} \\
&&+(k-1)(k-m)(k-1-m)\,d_{k-1}^{(m)}.
\end{eqnarray}
In particular, from the equation for $a_{k}^{(m)}$ at $k=0$, since $%
d_{1}^{(m)}=1$, we obtain

\begin{equation}
a_{0}^{(m)}=\frac{i}{\lambda }\,(m-1)\,.
\end{equation}

The introduced functions have simple analytic properties in the vicinity of
the point $x=0$. In particular, $\varphi _{r}^{(m)}(x,\lambda )$ is regular
here and is transformed under the modular transformation

\begin{equation}
x\rightarrow x^{\prime }=-x/(1-x)
\end{equation}
as

\begin{equation}
\varphi _{r}^{(m)}(x^{\prime },\lambda )=-(1-x)^{m}\,\varphi
_{r}^{(m)}(x,-\lambda )\,.
\end{equation}

The functions $\varphi _{s}^{(m)}(x,\lambda )$ and $\varphi
_{f}^{(m)}(x,\lambda )$ have singularities at $x=0$, which leads to
different results for their analytic continuations to negative values of $x$%
: 
\[
\varphi _{s}^{(m)}(x^{\prime },\lambda )=-(1-x)^{m}\,\left( \varphi
_{s}^{(m)}(x,-\lambda )\pm i\pi \,\varphi _{r}^{(m)}(x,-\lambda )\right) , 
\]
\begin{equation}
\varphi _{f}^{(m)}(x^{\prime },\lambda )=\exp (\pm i\pi
m)\,(1-x)^{m}\,\varphi _{f}^{(m)}(x,-\lambda ).
\end{equation}

Therefore, from the Bose symmetry of the Odderon wave function

\begin{equation}
f_{m,\widetilde{m}}(\overrightarrow{\rho _{1}},\overrightarrow{\rho _{2}},%
\overrightarrow{\rho _{3}};\overrightarrow{\rho _{0}})=\left( \frac{\rho
_{23}\,}{\,\rho _{20}\,\rho _{30}}\right) ^{m}\left( \frac{\,\rho _{23}^{*}\,%
}{\,\rho _{20}^{*}\,\rho _{30}^{*}}\right) ^{\widetilde{m}}\varphi _{m,%
\widetilde{m}}(x,x^{*})\,,\,
\end{equation}
combined with the single-valuedness condition near $\overrightarrow{x}=0$,
we obtain for the total wave function the following representation:

\[
\varphi _{m,\widetilde{m}}(x,x^{*})=\varphi _{f}^{(m)}(x,\lambda )\,\varphi
_{f}^{(\widetilde{m})}(x^{*},\lambda ^{*}) +c_{1}\left( \varphi
_{s}^{(m)}(x,\lambda )\,\varphi _{r}^{(\widetilde{m})}(x^{*},\lambda
^{*})+\varphi _{r}^{(m)}(x,\lambda )\,\varphi _{s}^{(\widetilde{m}%
)}(x^{*},\lambda ^{*})\right) 
\]
\begin{equation}
+c_{2} \,\varphi _{r}^{(m)}(x,\lambda )\,\varphi _{r}^{(\widetilde{m}%
)}(x^{*},\lambda ^{*})+\left( \lambda \rightarrow -\lambda \right) \, .
\end{equation}

The complex coefficients $c_{1},c_{2}$ and the eigenvalues $\lambda $ are
fixed from the conditions of the single-valuedness of $f_{m,\widetilde{m}}(%
\overrightarrow{\rho _{1}},\overrightarrow{\rho _{2}},\overrightarrow{\rho
_{3}};\overrightarrow{\rho _{0}})$ at $\overrightarrow{\rho _{3}}=%
\overrightarrow{\rho _{i}}$ ($i=1,2$) and the Bose symmetry [9]. It is
sufficient to require its invariance under the transformation $%
\overrightarrow{\rho _{2}}\leftrightarrow \overrightarrow{\rho _{3}}$
corresponding to the symmetry of $\varphi _{m,\widetilde{m}}(x,x^{*})$

\begin{equation}
\varphi _{m,\widetilde{m}}(x,x^{*})=\varphi _{m,\widetilde{m}%
}(1-x,1-x^{*})\,.
\end{equation}
For this purpose, one should analytically continue the functions $\varphi
_{i}^{(m)}(x)$ in the region near the point $x=1$ and calculate from the
differential equation the monodromy matrix $C_{rk}^{(m)}\,\,$ defined by the
relations 
\begin{eqnarray}
\varphi _{r}^{(m)}(x,\lambda ) &=&\sum_{k}C_{rk}^{(m)}\,\varphi
_{k}^{(m)}(1-x,-\lambda )\,,\, \\
\,\varphi _{r}^{(\widetilde{m})}(x^{*},\lambda ^{*}) &=&\sum_{k}C_{rk}^{(%
\widetilde{m})}\,\varphi _{k}^{(\widetilde{m})}(1-x^{*},-\lambda ^{*})\,.
\end{eqnarray}

Owing to the single-valuedness condition and the Bose symmetry of $f_{m,%
\widetilde{m}}$ we obtain a set of linear equations for parameters $c_{1}$
and $c_{2}$ with coefficients expressed in terms of $C_{rk}^{(m)}$ and $%
C_{rk}^{(\widetilde{m})}$. The spectrum of $\lambda $ is fixed by the
condition of selfconsistency of these equations [9].

To derive the relations among parameters $c_1, c_2$ and $\lambda$ following
from the duality symmetry, it is convenient to introduce the operators 
\begin{equation}
S_{m,\widetilde{m}}=S_m (x,p) \, S_{\widetilde{m}}(x^*,p^*)\,,\,\, S^{+}_{m,%
\widetilde{m}}=S^{t}_{1-m} (x,p) \, S^{t}_{1-\widetilde{m}}(x^*,p^*)\, ,
\end{equation}
where $S_m (x,p)$ and $S^{t}_{1-m}(x,p)$ were defined in the previous
section. These operators are hermitially conjugated each to another and have
the property: 
\begin{equation}
|A_m (x)|^2 = S^{+}_{m,\widetilde{m}}\,S_{m,\widetilde{m}}=S_{m,\widetilde{m}%
} \, S^{+}_{m,\widetilde{m}} \,.
\end{equation}
In particular it means, that they have common eigenfunctions 
\begin{equation}
S_{m,\widetilde{m}} \,\varphi _{m,\widetilde{m}}(x,x^{*})= \frac{|\lambda |^2%
}{c_1}\,e^{i\theta}\,\varphi _{m,\widetilde{m}}(x,x^{*})\,\,, \,\,\,S^{+}_{m,%
\widetilde{m}} \,\varphi _{m,\widetilde{m}}(x,x^{*})=
c_1\,e^{-i\theta}\,\varphi _{m,\widetilde{m}}(x,x^{*})\,,
\end{equation}
where 
\begin{equation}
e^{i\theta}=(-i)^{m-\widetilde{m}} \,\frac{\Gamma (m)}{\Gamma (1-\widetilde{m%
})} \frac{\Gamma (1+m) \Gamma (1+\widetilde{m})}{\Gamma (2-m) \Gamma (2-%
\widetilde{m})}\,.
\end{equation}
The eigenvalues are obtained from the small-$x$ asymptotics of these
equations.

Because the operators $S_{m,\widetilde{m}}$ and $S_{m,\widetilde{m}}^{+}$
are hermitially conjugated, we have 
\begin{equation}
|c_{1}|=|\lambda |.
\end{equation}
In the particular case $m=\widetilde{m}=1/2$, where $%
S_{1/2,1/2}=|S_{1/2}|^{2}$, the coefficient $c_{1}$ is positive: 
\begin{equation}
c_{1}=|\lambda |\,.
\end{equation}

Another relation can be derived if we shall take into account, that the
complex conjugated representations $\varphi _{m,\widetilde{m}}$ and $\varphi
_{1-m,1-\widetilde{m}}$ of the M\"obius group are related by the unitary
operator $U=R_m (x,p)R_{\widetilde{m}} (x^{*},p^{*})$, defined in the
previous section: 
\begin{equation}
e^{i\gamma}(\varphi _{m,\widetilde{m}})^{*}= U \,\varphi _{m,\widetilde{m}%
}\,,
\end{equation}
where $e^{i\gamma}$ is an eigenvalue of this operator. By calculating the
right hand side of this equation at $x \rightarrow 0$, we obtain 
\begin{equation}
e^{i\gamma}=(-1)^{m-\widetilde{m}}\, \frac{\Gamma (2-\widetilde{m}) \,\Gamma
(2-m)}{\Gamma (1+\widetilde{m})\, \Gamma (1+m)} \,\frac{c_1}{c_1^{*}}
\,\,,\,\,\, Im\,\frac{c_{2}}{c_{1}}= Im \,(m^{-1}+\widetilde{m}^{-1})\,.
\end{equation}

One can verify from the numerical results of ref. [9] that both relations
for $c_1$ and $c_2$ are fulfilled. For example, we have for the ground-state
eigenfunction with $m=\widetilde{m}=1/2$: 
\begin{equation}
i \,\lambda =0.205257506 \,\,,\,\,\,c_1=0.205257506
\end{equation}
and for one of the excited states with $m=\widetilde{m}=1/2+i \, 3/10$: 
\[
i \, \lambda = 0.247227544 \,,\,\,c_1=0.247186043-i \,0.004529717 \,,\,\,
c_2=-1.156524786-i \,0.415163678 , 
\]
\begin{equation}
|c_1|=|\lambda| \,\,,\,\,\, Im \,\frac{c_2}{c_1}= 2 \,Im \,(1/2+i \,
3/10)^{-1}\,.
\end{equation}

After the Fourier transformation of $\varphi _{m,\widetilde{m}}(%
\overrightarrow{x})$ to the momentum space $\overrightarrow{p}$ the regular
terms near the points $\overrightarrow{x}=0$ and $\overrightarrow{x}=1$ do
not give any contribution to its asymptotic behaviour at $\overrightarrow{p}%
\rightarrow \infty $. The requirement of the holomorphic factorization and
single-valuedness of the wave function in the momentum space leads to the
quantization of $\lambda $.

We can obtain from the duality equation and reality condition also the
representations for the coefficients $c_{1,2}$ in terms of integrals from $%
\varphi _{m,\widetilde{m}}(x,x^{*})$ over the fundamental region of the
modular group, where the expansion in $x$ is convergent. These relations
allow one to calculate the coefficients $c_{1,2}$ without using the
single-valuedness condition.

\section{Hamiltonian and integrals of motion}

The holomorphic Hamiltonian $h$ for the compound state of $n$ Reggeons for $%
N_{c}\rightarrow \infty $ commutes with the transfer matrix $T(u)$ owing to
the following relation for $h_{k,k+1}$: 
\begin{equation}
\left[ h_{k,k+1},T(u)\right] =-i\,tr\,\left( L_{1}(u)...L_{k-1}(u)\left(
L_{k}(u)-L_{k+1}(u)\right) L_{k+2}(u)...L_{n}(u)\right) .
\end{equation}
It can be considered as a linear equation for $h_{k,k+1}$. The formal
solution of this equation can be written as

\begin{equation}
h_{k,k+1}=\lim_{t\rightarrow \infty }\left( i\int_{0}^{t}d\,t\,^{^{\prime
}}\,\exp (i\,T(u)\,t^{\prime })\left[ h_{k,k+1},T(u)\right] \exp
(-i\,T(u)\,t^{\prime })+h_{k,k+1}(t)\right) \,,
\end{equation}
where $h_{k,k+1}(t)$ is the time-dependent operator

\begin{equation}
h_{k,k+1}(t)=\exp (i\,T(u)\,t)\,\,h_{k,k+1}\,\exp (-i\,T(u)\,t)\,\,.
\end{equation}

Since the integral term is cancelled in the sum of $h_{k,k+1}$, we can
substitute

\begin{equation}
h_{k,k+1}\rightarrow h_{k,k+1}(t).
\end{equation}
At $t\rightarrow \infty $ as a result of rapid oscillations of off-diagonal
matrix elements, each pair Hamiltonian is diagonalized in the
representation, where the transfer matrix is diagonal, and therefore it is a
function of the integrals of motion $\widehat{q_{k}}$:

\begin{equation}
h_{k,k+1}(\infty )=f_{k,k+1}(\widehat{q_{2}},\widehat{q_{3}},...\widehat{%
q_{n}})\,.
\end{equation}
Its dependence from the spectral parameter $u$ disappears in this limit and
the total Hamiltonian is

\begin{equation}
h=h(\widehat{q_2},\widehat{q_3},...\widehat{q_n})=\sum_{k=1}^nf_{k,k+1}(%
\widehat{q_2},\widehat{q_3},...\widehat{q_n}).
\end{equation}

All operators $O(t)$ satisfy the Heisenberg equations

\begin{equation}
-i\,\frac{d}{d\,t}O(t)=\left[ T(u)\,,\,O(t)\right] \,
\end{equation}
with certain initial conditions. In the case of the pair Hamiltonian the
initial conditions are

\begin{equation}
h_{k,k+1}(0)=\psi (\widehat{m}_{k,k+1})+\psi (1-\widehat{m}_{k,k+1})-2\psi
(1)\,,
\end{equation}
where the quantities $\widehat{m}_{k,k+1}$ are related to the pair Casimir
operators as

\begin{equation}
\widehat{m}_{k,k+1}(\widehat{m}_{k,k+1}-1)=M_{k,k+1}^{2}\,=-\rho
_{k,k+1}^{2}\,\partial _{k\,}\,\partial _{k+1}.
\end{equation}

In the case of the Odderon, $h$ does not depend on time if $h_{k,k+1}(t)$ is
determined as 
\begin{equation}
h_{k,k+1}(t)=e^{itA}\,h_{k,k+1}\,e^{-itA}\,\,,
\end{equation}
and $h_{k,k+1}(\infty )$ is a function of the total conformal momentum $%
\overrightarrow{M}^{2}=\widehat{m}(\widehat{m}-1)$ and of the integral of
motion $q_{3}=A$, which can be written as follows:

\begin{equation}
A=\frac{i^3}2\left[ M_{12}^2\,,\,M_{13}^2\right] =\frac{i^3}2\left[
M_{23}^2\,,\,M_{12}^2\right] =\frac{i^3}2\left[ M_{13}^2\,,\,M_{23}^2\right]
\,.
\end{equation}

Using these formulas and the following relations among the M\"obius group
generators $\overrightarrow{M}_r$

\begin{equation}
M_{ir}^{2}-M_{kr}^{2}=2\,\left( \overrightarrow{M}_{i}-\overrightarrow{M}%
_{k}\,,\,\overrightarrow{M}_{r}\right) \,,
\end{equation}
\begin{equation}
\left[ h_{ik},\left[ M_{ik}^{2}\,,\,\overrightarrow{M}_{i}-\overrightarrow{M}%
_{k}\right] \right] =4\left( \,\overrightarrow{M}_{i}-\overrightarrow{M}%
_{k}\right) \,,
\end{equation}
we can verify the commutation relations

\begin{equation}
i\left[ h_{12}\,,\,A\right] =M_{13}^{2}-M_{23}^{2}\,,\,\,i\left[
h_{13}\,,\,A\right] =M_{23}^{2}-M_{12}^{2},\,\,i\left[ h_{23}\,,\,A\right]
=M_{12}^{2}-M_{23}^{2}\,,
\end{equation}
from which it is obvious that $A$ commutes with $h$.

In a general case of $n$ reggeized gluons, one can use the Clebsch-Gordan
approach, based on the construction of common eigenfunctions of the total
momentum $\widehat{M}$ and a set $\left\{ \widehat{M}_{k}\right\} $ of the
commuting sub-momenta, to find all operators $M_{k,k+1}^{2}$ in the
corresponding representation. However to calculate $h$ we should perform an
unitary transformation to the representation, where $T(u)$ is diagonal,
because in this case for $t\rightarrow \infty $ the off-diagonal matrix
elements of $M_{k,k+1}^{2}$ disappear and their diagonal elements depend
only on $q_{r}$:

\begin{eqnarray*}
f_{k,k+1}(q_2,q_3,...q_n)=\langle q_2,...q_n\mid h_{k,k+1}\mid
q_2,...q_n\rangle \,,
\end{eqnarray*}
\begin{equation}
\widehat{q_k}\mid q_2,...q_n\rangle =q_k\mid q_2,...q_n\rangle \,.
\end{equation}

Let us consider, for example, the interaction between particles $1$ and $2$.
The transfer matrix, which should be diagonalized, can be written as follows

\begin{equation}
T(u)=\left( u^{2}-\frac{1}{2}\overrightarrow{L}^{2}\right)
d_{3...n}(u)+\left( i\,u\,\overrightarrow{L}-\frac{1}{4}\left[ 
\overrightarrow{L}^{2},\overrightarrow{N}\right] \right) \overrightarrow{d}%
_{3...n}(u)\,\,,
\end{equation}
where the differential operators $d_{3...n}(u)$ and $\overrightarrow{d}%
_{3...n}(u)$ are independent of $\overrightarrow{\rho _{1}}$ and $%
\overrightarrow{\rho _{2}}$. They are related to the monodromy matrix $%
t_{3...n}(u)\,$ for particles $3,4,...,n$ as follows

\begin{equation}
d_{3...n}(u)=tr\,t_{3...n}(u)\,,\,\overrightarrow{d}_{3...n}(u)=tr(%
\overrightarrow{\sigma }t_{3...n}(u)),\,\,t_{3...n}(u)=L_{3}(u)...L_{n}(u)\,
\end{equation}
and the matrix $t_{3...n}(u)$ satisfies the Yang-Baxter equations with a
hidden Lorentz symmetry.

The operators $\overrightarrow{L}$ and $\overrightarrow{N}$ are constructed
in terms of the M\"{o}bius group generators of particles $1$ and $2$:

\begin{equation}
\overrightarrow{L}=\overrightarrow{M}_1+\overrightarrow{M}_2\,,\,\,%
\overrightarrow{N}=\overrightarrow{M}_1-\overrightarrow{M}%
_2\,,\,\,M_k^z=\rho _k\partial _k\,,\,M_k^{+}=-\rho _k^2\partial
_k\,,\,M_k^{-}=\partial _k\,.
\end{equation}
They have the commutation relations, corresponding to the Lorentz algebra:

\[
\left[ L^{z},L^{\pm }\right] =\pm L^{\pm },\,\left[ L^{+},L^{-}\right]
=2L^{\pm },\,\left[ L^{z},N^{\pm }\right] =\pm N^{\pm },\, 
\]
\begin{equation}
\left[ L^{+},N^{-}\right] =2N^{z}\,,\,\left[ N^{z},N^{\pm }\right] =\pm
L^{\pm },\,\left[ N^{+},N^{-}\right] =2L^{z}\,.
\end{equation}

Let us introduce the Polyakov basis for the wave function of the composite
state of two gluons with the conformal weight $M$:

\begin{equation}
\mid \rho _{0^{\prime }},M\rangle =\left( \frac{\rho _{12}}{\rho
_{10^{\prime }}\rho _{20^{\prime }}}\right) ^{M}.
\end{equation}
Here $\rho _{0^{\prime }}$ enumerates the components of the
infinite-dimensional irreducible representation of the conformal group.

One can verify that the representation of the generators $\overrightarrow{L}$
and $\overrightarrow{N}$ in this basis is given by

\begin{eqnarray*}
M^z &\mid &\rho _{0^{\prime }},M\rangle =(\rho _1\partial _1+\rho _2\partial
_2)\mid \rho _{0^{\prime }},M\rangle =-(\rho _{0^{\prime }}\partial
_{0^{\prime }}+M)\mid \rho _{0^{\prime }},M\rangle , \\
M^{+} &\mid &\rho _{0^{\prime }},M\rangle =-(\rho _1^2\partial _1+\rho
_2^2\partial _2)\mid \rho _{0^{\prime }},M\rangle =(\rho _{0^{\prime
}}^2\partial _{0^{\prime }}+2M\rho _{0^{\prime }})\mid \rho _{0^{\prime
}},M\rangle ,
\end{eqnarray*}
\begin{equation}
M^{-} \mid \rho _{0^{\prime }},M\rangle =(\partial _1+\partial _2)\mid \rho
_{0^{\prime }},M\rangle =-\partial _{0^{\prime }}\mid \rho _{0^{\prime
}},M\rangle
\end{equation}
and

\begin{eqnarray*}
\left( N^{z}-\rho _{0^{\prime }}\,N^{-}\right) &\mid &\rho _{0^{\prime
}},M\rangle =\frac{M}{M-1}\,\,\partial _{0^{\prime }}\mid \rho _{0^{\prime
}},M-1\rangle , \\
\left( N^{+}+2\rho _{0^{\prime }}\,N^{z}-\rho _{0^{\prime }}^{2}N^{-}\right)
&\mid &\rho _{0^{\prime }},M\rangle =-2M\mid \rho _{0^{\prime }},M-1\rangle ,
\end{eqnarray*}
\begin{equation}
\frac{2M-1}{M(M-1)}N^{-}\mid \rho _{0^{\prime }},M\rangle =\,\mid \rho
_{0^{\prime }},M+1\rangle +\frac{1}{(M-1)^{2}}\,\,\partial _{0^{\prime
}}^{2}\mid \rho _{0^{\prime }},M-1\rangle .
\end{equation}

Note that there is a simple relation among the generators, provided that
they act on the state $\mid \rho _{0^{\prime }},M\rangle $:

\begin{equation}
\left[ N^{z}-\rho _{0^{\prime }}\,N^{-}\,,\,N^{+}+2\rho _{0^{\prime
}}\,N^{z}-\rho _{0^{\prime }}^{2}N^{-}\right] =M^{+}+2\rho _{0^{\prime
}}\,M^{z}-\rho _{0^{\prime }}^{2}M^{-}=0\,.
\end{equation}

The eigenfunction of the transfer matrix $T(u)$ can be written as a
superposition of the states $\mid \rho _{0^{\prime }},M\rangle $ with
various values of $\rho _{0^{\prime }}$ and $M$:

\begin{equation}
f_{m}(\rho _{1},\rho _{2},...,\rho _{n};\rho _{0})=\sum_{M}\int d\rho
_{0^{\prime }}\mid \rho _{0^{\prime }},M\rangle \,\,f_{m,M}(\rho _{0^{\prime
}},\rho _{3},...,\rho _{n};\rho _{0})\,,
\end{equation}
where $m$ is the conformal weight of the composite state. The function $%
f_{m,M}(\rho _{0^{\prime }}...\rho _{n};\rho _{0})$ in accordance with the
M\"{o}bius symmetry, has the form

\begin{equation}
f_{m,M}(\rho _{0^{\prime }},\rho _{3},...,\rho _{n};\rho _{0})=\left( \rho
_{0^{^{\prime }}0}\right) ^{-m+M-1}\prod_{r=3}^{n}\left( \frac{\rho _{r0}}{%
\rho _{r0^{^{\prime }}}}\right) ^{-\frac{m+M}{n-2}}\psi
(x_{1},x_{2},...,x_{n-3}),
\end{equation}
where $x_{r}$ are independent anharmonic ratios constructed from the
coordinates $\rho _{0^{\prime }},\rho _{3},...,\rho _{n}$.

Because of its M\"{o}bius invariance, the transfer matrix $T(u)$ after
acting on $f_{m}(\rho _{1},...,\rho _{n};\rho _{0})$ gives again a
superposition of the states $\mid \rho _{0^{\prime }},M\rangle $, but with
the coefficients which are linear combinations of $f_{m,M}$ and $f_{m,M\pm 1}
$. Therefore for its eigen function the coefficients satisfy some recurrence
relations, and the problem of the diagonalization of the transfer matrix $%
T(u)$ is reduced to the solution of these recurrence relations. For $n\geq 3$
in the sub-channel $\rho _{1,2}$ the recurrence relations depend on matrix
elements of the operators $\overrightarrow{d}_{3...n}(u)$ and $d_{3...n}(u)$
between the wave functions $f_{m,M}(\rho _{0^{\prime }},\rho _{3},...,\rho
_{n};\rho _{0})$ which should be chosen in such a way, to provide the
property of $f_{m}(\rho _{1},\rho _{2},...,\rho _{n};\rho _{0})$ to be a
representation of the cyclic group of transformations $i\rightarrow i+1$.

In the appendix we consider these relations in the first non-trivial case $%
n=3$. In the next section the relation between the Odderon Hamiltonian and
its integral of motion $A$ is discussed from another point of view.

\section{Odderon Hamiltonian in the normal order}

Let us write down the pair Hamiltonian as follows [4]

\begin{equation}
h_{12}=\log (\rho _{12}^{2}\,\partial _{1})+\log (\rho _{12}^{2}\,\partial
_{2})-2\,\log (\rho _{12}\,)-2\,\psi (1)\,.
\end{equation}
This representation allows us to present the total Hamiltonian for $n$
reggeized gluons in the form invariant under the M\"{o}bius transformations

\begin{equation}
h=\sum_{k=1}^{n}\left( \log \left( \frac{\rho _{k+2,0}\,\,\rho _{k,k+1}^{2}}{%
\rho _{k+1,0}\,\,\rho _{k+1,k+2}}\,\partial _{k}\right) +\log \left( \frac{%
\rho _{k-2,0}\,\,\rho _{k,k-1}^{2}}{\rho _{k-1,0}\,\,\rho _{k-1,k-2}}%
\,\partial _{k}\right) -2\,\psi (1)\right) \,,
\end{equation}
where $\rho _{0}$ is the coordinate of the composite state.

We consider below in more detail the Odderon. Using for its wave function
the conformal anzatz

\begin{equation}
f_{m}(\rho _{1},\rho _{2},\rho _{3};\rho _{0})=\left( \frac{\rho _{23}}{\rho
_{20}\rho _{30}}\right) ^{m}\varphi _{m}(x)\,,\,\,\,\,x=\frac{\rho _{12}\rho
_{30}}{\rho _{10}\rho _{32}}\,,\,\,
\end{equation}
one can obtain the following Hamiltonian for the function $\varphi _{m}(x)$
in the space of the anharmonic ratio $x$ [4]

\[
h=6\gamma +\log \left( x^{2}\partial \right) +\log \left( (1-x)^{2}\partial
\right) +\log \left( \frac{x^{2}}{1-x}((1-x)\partial +m)\right) + 
\]

\begin{equation}
\log \left( \frac{1}{1-x}((1-x)\partial +m))\right) +\,\log \left( \frac{%
(1-x)^{2}}{x}(x\partial -m)\right) +\log \left( \frac{1}{x}(x\partial
-m)\right) .
\end{equation}

It is convenient to introduce the logarithmic derivative $P\equiv x\partial $
as a new momentum. Using the relations [4]:

\begin{eqnarray*}
\log (x^{2}\partial ) &=&\log (x)+\psi (1-P)\,,\,\,\,\,\log (\partial
)=-\log (x)+\psi (-P)\,\,,\,\, \\
\log (x^{2}\partial ) &=&\log (\partial )+2\log (x)-\frac{1}{P}\,,
\end{eqnarray*}
\begin{equation}
((1-x)\partial +m)=(1-x)^{1+m}\,\partial \,(1-x)^{-m}\,,\,\,\,\,\,x\partial
-m=x^{1+m}\partial \,x^{-m}\,,
\end{equation}
one can transform this Hamiltonian to the normal form:

\begin{equation}
\frac{h}{2}=\,\,-\log (x)+\psi (1-P)+\psi (-P)+\psi (m-P)-3\psi
(1)+\sum_{k=1}^{\infty }x^{k}\,f_{k}(P)\,,
\end{equation}
where

\begin{equation}
f_{k}(P)=-\frac{2}{k}+\frac{1}{2}\left( \frac{1}{P+k-m}+\frac{1}{P+k}\right)
+\sum_{t=0}^{k}\frac{c_{t}(k)}{P+t}\,.
\end{equation}
Here 
\begin{equation}
c_{t}(k)=\frac{(-1)^{k-t}\,\,\Gamma (m+t)\,\left( (t-k)\,(m+t)+m\,k/2\right) 
}{k\,\Gamma (m-k+t+1)\,\Gamma (t+1)\,\Gamma (k-t+1)}\,.
\end{equation}

On the other hand the differential operators $a_{m}$ and $a_{1-m}$ can be
written in terms of the quantities $P$ and $x$ as follows

\begin{eqnarray}
a_{m} &=&i^{-1-m}x^{-m}(1-x)\frac{\Gamma (m-P+1)}{\Gamma (-P)}\,,\,\, \\
a_{1-m} &=&i^{-2+m}x^{-1+m}(1-x)\frac{\Gamma (-P-m+2)}{\Gamma (-P)}\,.
\end{eqnarray}

Using the above representation for $h$ and the following expression for the
integral of motion:

\begin{equation}
B=i\,a_{1-m}\,a_{m}=\frac{(1-x)}{x}\left( (1-x)P-1-x+xm\right) P\,(P-m) \,,
\end{equation}
one can verify their commutativity 
\begin{equation}
\left[ h,B\right] =0\,.
\end{equation}
Therefore $h$ is a function of $B$.

In particular for large $B$ this function should have the form:

\begin{equation}
\frac{h}{2}=\,\,\log (B)+3\gamma +\sum_{r=1}^{\infty }\frac{c_{r}}{B^{2r}}\,.
\end{equation}
The first two terms of this asymptotic expansion were calculated in ref.
[4]. The series is constructed in inverse powers of $B^{2}$, because $h$
should be invariant under all modular transformations, including the
inversion $x\rightarrow 1/x$ under which $B$ changes its sign. The same
functional relation should be valid for the eigenvalues $\varepsilon /2$ and 
$\mu =i\,\lambda $ of these operators: 
\begin{equation}
\frac{\varepsilon }{2}=\,\,\log (\mu )+3\gamma +\sum_{r=1}^{\infty }\frac{%
c_{r}}{\mu ^{2r}}\,.
\end{equation}

For large $\mu $ it is convenient to consider the corresponding eigenvalue
equations in the $P$ representation, where $x$ is the shift operator 
\begin{equation}
x=\exp (-\frac{d}{dP})\,,
\end{equation}
after extracting from eigenfunctions of $B$ and $h$ the common factor 
\begin{equation}
\varphi _{m}(P)=\Gamma (-P)\,\Gamma (1-P)\,\Gamma (m-P)\,\exp (i\pi
P)\,\,\Phi _{m}(P).
\end{equation}

The function $\Phi _{m}(P)$ can be expanded in series over $1/\mu $ 
\begin{equation}
\Phi _{m}(P)=\sum_{n=0}^{\infty }\mu ^{-n}\Phi _{m}^{n}(P)\,,\,\Phi
_{m}^{0}(P)=1\,,
\end{equation}
where the coefficients $\Phi _{m}^{n}(P)$ turn out to be the polynomials of
order $4n$ satisfying the recurrence relation: 
\[
\Phi _{m}^{n}(P)=\sum_{k=1}^{P}(k-1)\,(k-1-m)\,\left( (k-m)\,\Phi
_{m}^{n-1}(k-1)+(k-2)\,\Phi _{1-m}^{n-1}(k-1-m)\right) 
\]
\begin{equation}
-\frac{1}{2}\sum_{k=1}^{m}(k-1)\,(k-1-m)\,\left( (k-m)\,\Phi
_{m}^{n-1}(k-1)+(k-2)\,\Phi _{1-m}^{n-1}(k-1-m)\right) ,
\end{equation}
valid due to the duality equation written below for a definite choice of the
phase of $\Phi _{m}(P)$  

\begin{equation}
x^{-m}\left( 1-x\,P\,(P-m)\,(P-m+1)\right) \,\Phi _{m}(P)=\mu ^{m}\,\Phi
_{1-m}(P)\,
\end{equation}
with the use of the substitution $x\,\mu \rightarrow x$.

Note that the summation constants $\Phi _{m}^{n}(0)$ in the above recurrence
relation have the anti-symmetry property 
\begin{equation}
\Phi _{m}^{n}(0)=-\Phi _{1-m}^{n}(0)\,,
\end{equation}
which guarantees the fulfilment of the relation 
\begin{equation}
\Phi _{m}^{n}(m)=\Phi _{1-m}^{n}(0\,)\,
\end{equation}
being a consequence of the duality relation

\begin{equation}
\Phi _{m}^{n}(P)=\Phi _{1-m}^{n}(P-m)+(P-1)(P-m)(P-m-1)\Phi
_{m}^{n-1}(P-1)\,.
\end{equation}
The symmetric part of $\Phi _{m}^{n}(0)$ under the substitution $%
m\leftrightarrow 1-m$ would simply modify the normalization constant for $%
\Phi _{m}(P)$.

The energy can be expressed in terms of $\Phi _{m}(P)$ as follows:

\[
\frac{\varepsilon }{2}=\,\,\log (\mu )+3\gamma +\frac{\partial}{\partial P}%
\log \Phi _{m}(P) 
\]
\begin{equation}
+\left( \Phi _{m}(P)\right)^{-1} \, \sum_{k=1}^{\infty }\mu
^{-k}\,f_{k}(P-k)\,\Phi _{m}(P-k)\prod_{r=1}^{k}(P-r)(P-r+1)(P-r-m+1)\,
\end{equation}
and it should not depend on $P$ due to the commutativity of $h$ and $B$.

By solving the recurrence relations for $\Phi _{m}^{n}(P)$ and putting the
result in the above expression, we obtain the following asymptotic expansion
for $\varepsilon /2$:

\[
\frac{\varepsilon }{2}=\log (\mu )+3\gamma +\left( \frac{3}{448}+\frac{13}{%
120}(m-1/2)^{2}-\frac{1}{12}(m-1/2)^{4}\right) \frac{1}{\mu ^{2}}+ 
\]

\begin{equation}
\left( -\frac{4185}{2050048}-\frac{2151}{49280}(m-1/2)^{2}+...\right) \frac{1%
}{\mu ^{4}}+\left( \frac{965925}{37044224}+...\right) \frac{1}{\mu ^{6}}%
+...\,\,\,.
\end{equation}
This expansion can be used with a certain accuracy even for the smallest
eigenvalue $\mu =0.20526$, corresponding to the ground-state energy $%
\varepsilon =0.49434$ [9]. For the first excited state with the same
conformal weight $m=1/2$, where $\varepsilon = 5.16930$ and $\mu = 2.34392$
[9], the energy can be calculated from the above asymptotic series with a
good precision. The analytic approach, developed in this section, should be
compared with the method based on the Baxter equation [10].

In the conclusion, we note that the remarkable properties of the Reggeon
dynamics are presumably related with supersymmetry. In the continuum limit $%
n\rightarrow \infty $ the above duality transformation coincides with the
supersymmetric translation, which is presumably connected with the
observation [11], that in this limit the underlying model is a twisted $N=2$
supersymmetric topological field theory. Additional arguments supporting the
supersymmetric nature of the integrability of the reggeon dynamics were
given in ref. [12]. Namely, the eigenvalues of the integral kernels in the
evolution equations for quasi-partonic operators in the $N=4$ supersymmetric
Yang-Mills theory are proportional to $\psi (j-1)$, which means that these
evolution equations in the multicolour limit are equivalent to the
Schr\"{o}dinger equation for the integrable Heisenberg spin model similar to
the one found in the Regge limit [7]. Note that at large $N_{c}$ the $N=4$
Yang-Mills theory is guessed to be related with the low-energy asymptotics
of a superstring model [13].

\[
\]
{\large {\bf Acknowledgements}}\newline

I want to thank the Universit\"{a}t Hamburg for its hospitality during my
stay in Germany, where the basic part of this work was done. I thank G.
Altarelli, J. Ellis and other participants of the CERN theory seminar for
their interest in my talk. Fruitful discussions with L. Faddeev, A. Neveu,
V. Fateev, A. Zamolodchikov, J. Bartels, A. Martin, B. Nicolescu, P. Gauron,
E. Antonov, M. Braun, A. Bukhvostov, S. Derkachev, A. Manashov, G. Volkov,
R. Kirschner, L. Szymanowski and J. Wosiek were helpful.

\vspace{1cm} \noindent 
\[
\]

\section*{Appendix}

Here we consider consequences of the conformal weight representation for the
Odderon. In this case the total Hamiltonian is

\[
h(m,\lambda )\,=\,h_{12}+h_{23}+h_{31}\,. 
\]

The eigenvalue of $h$ is expressed in terms of its matrix elements:

\[
h(m,\lambda )=\sum_{k=1}^{3}\langle m,\lambda \mid h_{k,k+1}\mid m,\lambda
\rangle \,, 
\]
where $\mid m,\lambda \rangle $ is a normalized eigenfunction of two
commuting operators

\[
\left( \sum_{k=1}^{3}\overrightarrow{M_{k}}\right) ^{2}\mid m,\lambda
\rangle =m(m-1)\mid m,\lambda \rangle \,,\,\,\,\,A\mid m,\lambda \rangle
=\lambda \mid m,\lambda \rangle \,. 
\]

Let us consider for definiteness the interaction in the channel $12$, where $%
M$ is the pair conformal weight. If one will construct the matrix $%
V_{M}^{\lambda }(m)$ performing the unitary transformation between the $M$-
and $\lambda $-representations

\[
\mid m,\lambda \rangle =\sum_{M}V_{M}^{\lambda }(m)\mid m,M\rangle
\,,\,\,\,M_{12}^{2}\mid m,M\rangle =M(M-1)\mid m,M\rangle
\,,\,\,\,\sum_{M}V_{M}^{\lambda }\,V_{\lambda ^{\prime }}^{M}=\delta
_{\lambda ^{\prime }}^{\lambda }, 
\]
then the diagonal matrix elements of the pair Hamiltonian $h_{12}$ can be
calculated as

\[
\langle m,\lambda \mid h_{12}\mid m,\lambda \rangle
=\sum_{M}h(M)\,V_{M}^{\lambda }\,(m)V_{\lambda }^{M}(m)\,,\,\,h(M)=\psi
(M)+\psi (1-M)+2\gamma \,. 
\]

We shall derive below the recurrence relations for $V_{M}^{\lambda }$. To
begin with, we note that, according to the commutation relations for the
Lorentz algebra generators, there are non-vanishing matrix elements of the
boost operator $\overrightarrow{N}$ only between the states $\mid M\rangle $
and $\mid M\pm 1\rangle $. This is valid also for the matrix elements of the
operators $A$ and $M_{13}^{2}-M_{23}^{2}$ according to the relations

\[
\langle M^{\prime }\mid A\mid M\rangle =\frac{i^{3}}{4}(M^{\prime
}-M)\,(M^{\prime }+M-1)\,\,\langle M^{\prime }\mid M_{13}^{2}-M_{23}^{2}\mid
M\rangle \,, 
\]
\[
\langle M\pm 1\mid M_{13}^{2}-M_{23}^{2}\mid M\rangle =2\overrightarrow{M_{3}%
}\,\langle M\pm 1\mid \overrightarrow{N}\mid M\rangle . 
\]

Thus, for the common eigenfunctions $\mid m,M\rangle $ of two commuting
Casimir operators

\[
\left( \sum_{k=1}^{3}\overrightarrow{M_{k}}\right) ^{2}\mid m,M\rangle
=m(m-1)\mid m,M\rangle \, 
\]
and 
\[
M_{12}^{2}\mid m,M\rangle =M(M-1)\mid m,M\rangle \, 
\]
we have

\[
A\mid m,M\rangle =\frac{i^3}2\left( MC_{m,M}^{+}\mid m,M+1\rangle
-(M-1)C_{m,M}^{-}\mid m,M-1\rangle \right) \,, 
\]
where the coefficients $C_{m,M}^{\pm }$ are defined by the relations

\[
\left( M_{13}^2-M_{23}^2\right) \mid m,M\rangle =C_{m,M}^{+}\mid
m,M+1\rangle +C_{m,M}^{-}\mid m,M-1\rangle \,. 
\]

One can obtain from the above equations the following recurrence relation
for the unitary matrix $V_{M}^{\lambda }(m)$:

\[
\lambda \,V_M^\lambda (m)=\frac{i^3}2\left(
(M-1)C_{m,M-1}^{+}\,V_{M-1}^\lambda (m)-MC_{m,M+1}^{-}V_{M+1}^\lambda
(m)\right) . 
\]

To calculate the matrix elements $C_{m,M}^{\pm }$ of the operator $%
M_{13}^{2}-M_{23}^{2}$, we use the above representation of the generators $%
\overrightarrow{L}$ and $\overrightarrow{N}$ in the Polyakov basis and obtain

\[
\left( M_{13}^{2}-M_{23}^{2}\right) \mid \rho _{0^{\prime }},M\rangle
=\left( 2N^{z}M_{3}^{z}+N^{+}M_{3}^{-}+N^{-}M_{3}^{+}\right) \,\mid \rho
_{0^{\prime }},M\rangle = 
\]
\[
2M\left( \frac{1}{M-1}\,\rho _{30^{\prime }}\partial _{0^{\prime }}-1\right)
\mid \rho _{0^{\prime }},M-1\rangle \,\partial _{3}-\rho _{30^{\prime
}}^{2}\,N^{-}\mid \rho _{0^{\prime }},M\rangle \,\partial _{3}\,. 
\]

Owing to the M\"{o}bius invariance, the three-gluon state $\mid m,M\rangle $%
, with the conformal weights $m$ and $M$, can be written as a superposition
of the Polyakov functions

\[
f_{m,M}(\rho _{1},\rho _{2},\rho _{3};\rho _{0})=\int_{L}d\rho _{0^{\prime
}}\,\mid \rho _{0^{\prime }},M\rangle \,\left( \rho _{0^{^{\prime
}}0}\right) ^{-m+M-1}\left( \frac{\rho _{30}}{\rho _{30^{^{\prime }}}}%
\right) ^{-M-m+1}
\]
with various integration contours $L$. By integrating the terms in $M_{3}^{i}
$ with derivatives of $\mid \rho _{0^{\prime }},M\rangle $ by parts and
using the relations

\[
-\rho _{30^{\prime }}^{2}\partial _{3}\left( \rho _{0^{^{\prime }}0}\right)
^{-m+M-1}\left( \frac{\rho _{30}}{\rho _{30^{^{\prime }}}}\right)
^{-M-m+1}=-(M+m-1)\,\left( \rho _{0^{^{\prime }}0}\right) ^{-m+M}\,\left( 
\frac{\rho _{30}}{\rho _{30^{^{\prime }}}}\right) ^{-M-m},
\]
\[
\left( \frac{1}{1-2M}\,\partial _{0^{\prime }}^{2}\,\rho _{30^{\prime
}}^{2}\,-2\,\partial _{0^{\prime }}\,\rho _{30^{\prime }}-2(M-1)\right)
\,\partial _{3}\left( \rho _{0^{^{\prime }}0}\right) ^{-m+M-1}\left( \frac{%
\rho _{30}}{\rho _{30^{^{\prime }}}}\right) ^{-M-m+1}=
\]
\[
-(M+m-1)\,\frac{(m-M)(m-M+1)}{2M-1}\,\left( \rho _{0^{^{\prime }}0}\right)
^{-m+M-2}\left( \frac{\rho _{30}}{\rho _{30^{^{\prime }}}}\right) ^{-M-m+2},
\]
one can obtain the recurrence relation for the function $f_{m,M}=\,f_{m,M}(%
\rho _{1},\rho _{2},\rho _{3};\rho _{0})$: 
\[
\left( M_{13}^{2}-M_{23}^{2}\right) \,f_{m,M}=\frac{M(M+m-1)}{1-2M}\left(
(M-1)\,f_{m,M+1}+\frac{(m-M)(m-M+1)}{M-1}f_{m,M-1}\right) .
\]

Due to its M\"{o}bius covariance  $f_{m,M}(\rho _{1},\rho _{2},\rho
_{3};\rho _{0})$ can be presented in the form

\[
f_{m,M}(\rho _{1},\rho _{2},\rho _{3};\rho _{0})=\left( \frac{\rho _{12}\rho
_{23}\rho _{31}}{\rho _{10}^{2}\rho _{20}^{2}\rho _{30}^{2}}\right)
^{m/3}f_{m,M}(x)\,, 
\]
where $x$ is the anharmonic ratio 
\[
x=\frac{\rho _{12}\,\rho _{30}}{\rho _{10}\,\rho _{32}}\,.\, 
\]

By introducing the new integration variable 
\[
x^{^{\prime }}=\frac{\rho _{12}\rho _{30^{^{\prime }}}}{\rho _{10^{^{\prime
}}}\rho _{32}}\, 
\]
we obtain the following expression for $f_{m,M}(x)\,$:

\[
f_{m,M}(x)=\left( x(1-x)\right) ^{2m/3}\int dx^{^{\prime }}(1-x^{^{\prime
}})^{-M}(x^{^{\prime }}-x)^{-m+M-1}\left( \frac{x}{x^{^{\prime }}}\right)
^{-M-m+1}. 
\]

This function satisfies the following differential equation

\[
M_{12}^{2}(x)\,f_{m,M}(x)=M(M-1)\,f_{m,M}(x)\, 
\]
and the recurrence relation

\[
\left( M_{13}^{2}(x)-M_{23}^{2}(x)\right) \,f_{m,M}(x)= 
\]
\[
\frac{M(M+m-1)}{1-2M}\left( (M-1)\,f_{m,M+1}(x)+\frac{(m-M)(m-M+1)}{M-1}%
f_{m,M-1}(x)\right) \,. 
\]
The pair Casimir operators in the $x$-representation are given below

\[
M_{12}^{2}(x)=\frac{x}{1-x}\left( \frac{m}{3}(1-2x)+x(1-x)\partial \right) 
\frac{1}{x}\left( \frac{m}{3}(1+x)+x(1-x)\partial \right) , 
\]
\[
M_{13}^{2}(x)=\frac{1-x}{x}\left( \frac{m}{3}(1-2x)+x(1-x)\partial \right) 
\frac{1}{1-x}\left( \frac{m}{3}(x-2)+x(1-x)\partial \right) , 
\]
\[
M_{23}^{2}(x)=-\frac{1}{x(1-x)}\left( \frac{m}{3}(x-2)+x(1-x)\partial
\right) \left( \frac{m}{3}(1+x)+x(1-x)\partial \right) 
\]
and satisfy the relation:

\[
M_{12}^{2}(x)+M_{13}^{2}(x)+M_{23}^{2}(x)=m(m-1)\,. 
\]

The function $f_{m,M}(x)$ can be expressed for two different choices of the
integration contour $L$ through the hypergeometric functions:

\[
f_{m,M}^1(x)=\frac{\Gamma (1-M)}{\Gamma (M)\Gamma (2-2M)}\,\left(
x(1-x)\right) ^{2m/3}x^{1-M-m}F(m+1-M,1-M;2-2M;x)\, , 
\]
\[
f_{m,M}^2(x)=\frac{\Gamma (m+M)}{\Gamma (2M)\Gamma (1+m-M)}\left(
x(1-x)\right) ^{2m/3}x^{M-m}F(m+M,M;2M;x)\,, 
\]
where

\[
F(a,b;c;x)=1+\frac{ab}{c}\frac{x}{1!}+\frac{a(a+1)b(b+1)}{c(c+1)}\frac{x^{2}%
}{2!}+...\,\,. 
\]

Moreover, the above eigenvalue equation for $f_{m,M}(x)$ is equivalent to
the hypergeometric equation for $F(a,b;c;x)$:

\[
x(1-x)\frac{d^2}{dx^2}F+\left( c-(a+b+1)x\right) \frac d{dx}F-abF=0\,, 
\]
because 
\[
\left( x(1-x)\right) ^{-2m/3}x^{-M+m}M_{12}^2(x)\left( x(1-x)\right)
^{2m/3}x^{M-m}= M(M-1)+ 
\]
\[
x\,\left( x(1-x)\frac{d^2}{dx^2}+\left( 2M-(m+2M+1)x\right) \frac
d{dx}-M(M+m)\right) \,. 
\]

In an analogous way, using the relation

\[
\left( x(1-x)\right) ^{-2m/3}x^{-M+m}\left(
M_{13}^{2}(x)-M_{23}^{2}(x)\right) \left( x(1-x)\right) ^{2m/3}x^{M-m}= 
\]
\[
\frac{1}{x}(M(2-x)+(1-m)x+(2-x)x\partial )\left( M(1-x)-m+x(1-x)\partial
\right) 
\]
and extracting from it the hypergeometric differential operator with the
coefficient chosen to cancel the term with the second derivative, we obtain
that the above recurrence relation for $f_{m,M}(x)$ is equivalent to the
following relation for the hypergeometric function $F_{M}(x)=F(M,M+m;2M;x)$: 
\[
\left( m(m-1)+M(2m-1-M)+2(1-m)(1-x)\partial +\frac{2M(M-m)}{x}\right)
F_{M}(x) 
\]
\[
=(M-m)\left( \frac{(M-1)(M+m)(M+m-1)}{2(2M-1)(2M+1)}\,x\,F_{M+1}(x)+\frac{2M%
}{x}F_{M-1}(x)\right) . 
\]
One can verify its validity in the $k$-th order of the expansion in $x$ by
taking into account the algebraic identity:

\[
(m+2k)(m-1)+M(2m-1-M)+2\left( 1-m+\frac{M(M-m)}{k+1}\right) \frac{%
(M+k)(M+m+k)}{2M+k} 
\]
\[
=\frac{(M-m)(M+m-1)}{2M-1}\left( \frac{(M-1)k}{2M+k}+\frac{M(2M+k-1)}{k+1}%
\right) \,. 
\]

Let us now return to the problem of finding the recurrence relations for the
matrix $V^{\lambda }_M (m)$ performing the unitary transformation between
the $M$- and $\lambda $-representations. As earlier, it is convenient to
work with the function $\varphi _m(x)$ defined by the relation 
\[
f_m (x) =(x(1-x))^{-m/3} \,\varphi _m (x) \,. 
\]
It satisfies the equation 
\[
a_{1-m}\,a_m \varphi _m (x)=\lambda \varphi _m (x)\,, 
\]
where 
\[
a_{m}=x(1-x)(i\partial )^{m+1}. 
\]

One can search the solution of the above eigenvalue equation as the linear
combination 
\[
\varphi _m(x)=\sum_{M}C_{M}(m)\,F_{M}^{m}(x)\, 
\]
of the eigenfunctions of the operator $M_{12}^{2}$. The coefficients $C_M
(m) $ would coincide with the matrix $V^{\lambda }_M (m)$, if the functions $%
F_{M}^{m} (x)$ would be normalized. However, in accordance with above
formulas we define $F_{M}^{m}(x)$ by the following expression: 
\[
F_{M}^{m}(x)=\frac{\Gamma (2M)\Gamma (1+m-M)}{\Gamma (m+M)}\left(
x(1-x)\right) ^{m/3}f_{m,M}^{2}(x)\,=x^{M}(1-x)^{m}\,F(m+M,M;2M;x)\,. 
\]
The transition to the renormalized functions can be easily done.

According to the above presentation the quantities $F_M^m(x)$ satisfy the
following equality:

\[
a_{1-m}a_{m}\,F_{M}^{m}\,(x)=\frac{i^{3}}{2}M(M-1)(M-m)\left( \frac{%
(M+m)(M+m-1)}{2(2M-1)(2M+1)}\,F_{M+1}^{m}(x)-2F_{M-1}^{m}(x)\right) .
\]
Because the right-hand side of this equality is zero for $M=0,1$ and $m$,
one can restrict the summation over $M$ in the eigenfunction $\varphi _{m}(x)
$ to two series: $M=r$ and $M=m+r$ with $r=0,1,2,...$. However, in the first
case the coefficient in front of $F_{1}^{m}(x)$ can not be calculated
through the coefficient in front of $F_{0}^{m}(x)$. Due to this degeneracy
of the equation one should introduce for $r\geq 1$ the more complicated
function $\Phi _{r}^{m}(x)$: 
\[
\Phi _{r}^{m}(x)=\lim_{M\rightarrow r}\frac{d}{dM}F_{M}^{m}(x)\,.
\]
The recurrence relation for these functions can be obtained by
differentiating the relation for $F_{M}^{m}(x)$ . In particular, for $r=1$
we obtain

\[
a_{1-m}a_{m}\,\Phi _{1}^{m}\,(x)=\frac{i^{3}}{2}(1-m)\left( \frac{(1+m)\,m}{6%
}\,F_{2}^{m}(x)-2F_{0}^{m}(x)\right) \,. 
\]

Thus, in accordance with the small-$x$ behaviour of the eigenfunctions $%
\varphi (x)$, discussed above, we write the linearly independent solutions
in the form: 
\[
\varphi _{r}^{(m)}(x,\lambda )=\sum_{k=1}^{\infty }\Delta _{k}^{m}(\lambda
)\,F_{k}^{m}(x)\,.
\]
\[
\varphi _{s}^{(m)}(x,\lambda )=\sum_{k=0}^{\infty }\left( \alpha
_{k}^{m}(\lambda )\,F_{k}^{m}(x)+\Delta _{k}^{m}(\lambda )\Phi
_{k}^{m}\,(x)\right) \,,
\]
\[
\varphi _{f}^{(m)}(x,\lambda )=\sum_{k=0}^{\infty }\gamma _{k}^{m}(\lambda
)\,F_{k+m}^{m}(x)\,,
\]
The coefficients $\alpha _{k}^{m},\gamma _{k}^{m}$ and $\Delta _{k}^{m}$
satisfy the recurrence relations: 
\[
i\lambda \,\alpha _{k}^{m}(\lambda )=\left( \alpha _{k+1}^{m}(\lambda
)+\beta _{k+1}^{m}(\lambda )\,\frac{d}{dk}\right) \,k(k+1)(k-m+1)\,
\]
\[
-\frac{1}{4}\left( \alpha _{k-1}^{m}(\lambda )+\beta _{k-1}^{m}(\lambda )\,%
\frac{d}{dk}\right) (k-1)(k-2)(k-m-1)\frac{(k+m-1)(k+m-2)}{(2k-3)(2k-1)}\,,
\]
\[
i\lambda \,\Delta _{k}^{m}(\lambda )=-\frac{1}{4}(k-1)(k-2)(k-m-1)\frac{%
(k+m-1)(k+m-2)}{(2k-3)(2k-1)}\Delta _{k-1}^{m}(\lambda )
\]
\[
+k(k+1)(k-m+1)\,\Delta _{k+1}^{m}(\lambda )\,,\,\,\,\Delta _{1}^{m}(\lambda
)=1\,;
\]
\[
i\lambda \,\gamma _{k}^{m}(\lambda )=-\frac{1}{4}(k+m-1)(k+m-2)(k-1)\frac{%
(k+2m-1)(k+2m-2)}{(2k+2m-3)(2k+2m-1)}\gamma _{k-1}^{m}(\lambda )
\]
\[
+(k+m)(k+m+1)(k+1)\,\gamma _{k+1}^{m}(\lambda )\,,\,\,\,\gamma
_{0}^{m}(\lambda )=1\,.
\]

If we compare these relations with the derived above analogous recurrence
relations for the coefficients of the expansion of $\varphi _{r}^{(m)}(x)$
in the series over $x$, it is obvious that the factors in front of the
corresponding quantities with the index $k+1$ coincide. Furthermore, the
quantities $\alpha _{k},\,\gamma _{k}$ and $\Delta _{k}$ are absent in the
right-hand side of these relations contrary to the previous case, where the
similar factors in front of $a_{k}\,,\,c_{k}$ and $d_{k}$ are non-zero.

\[
\]

\section*{References}

\[
\]

1. L.N. Lipatov, Sov. J. Nucl. Phys. ${\bf 23}$ (1976) 642;

V.S. Fadin, E.A. Kuraev and L.N. Lipatov, Phys. Lett. ${\bf B60}$ (1975) 50;

Ya.Ya. Balitsky and L.N. Lipatov, Sov. J. Nucl. Phys. ${\bf 28}$ (1978) 822;

L.N. Lipatov, Sov. Phys. JETP ${\bf 63}$ (1986) 904;

2. V.S. Fadin, L.N. Lipatov, Phys. Lett. ${\bf {B429}}$ (1998) 127;

3. J.Bartels, Nucl. Phys. ${\bf {B175}}$ (1980) 365 ;

J. Kwiecinski and M. Prascalowicz, Phys. Lett. ${\bf {B94}}$ (1980) 413;

4. L.N. Lipatov, Phys. Lett. ${\bf {B251}}$ (1990) 284; ${\bf {B309}}$
(1993) 394

5. L.N. Lipatov, hep-th/9311037, Padua preprint DFPD/93/TH/70, unpublished;

6. R.J. Baxter, Exactly Solved Models in Statistical Mechanics, (Academic
Press,

New York, 1982);

V.O. Tarasov, L.A. Takhtajan and L.D. Faddeev, Theor. Math. Phys. ${\bf {57}}
$ (1983) 163;

7. L.N. Lipatov, Sov. Phys. JETP Lett. ${\bf {59}}$ (1994) 571;

L.D. Faddeev and G.P. Korchemsky, Phys. Lett. ${\bf {B342}}$ (1995) 311;

8. C. Montonen and D. Olive, Phys. Lett. ${\bf 72}$ (1977) 117;

N. Seiberg and E. Witten, Nucl. Phys. ${\bf 426}$ (1994) 19;

9. L.N. Lipatov, Recent Advances in Hadronic Physics, Proceedings of the
Blois

conference (World Scientific, Singapore, 1997);

R.Janik and J. Wosiek, hep-ph/9802100, Crakow preprint TPJU-2/98;

M.A. Braun, hep-ph/9801352, St.Petersburg University preprints;

M.A. Braun, P. Gauron and B. Nicolescu, preprint LPTPE/UP6/10/July 98;

M. Praszalowicz and A. Rostworowski, hep-ph/9805245, Crakow preprint
TPJU-8/98;

10. R. Janik and J. Wosiek, Phys. Rev. Lett. ${\bf 79}$ (1997) 2935;

11. J. Ellis and N.E. Mavromatos, preprint OUTP-98-51P, hep-ph/9807451;

12. L.N. Lipatov, Perspectives in Hadronics Physics, Proceedings of the ICTP

conference (World Scientific, Singapore, 1997).

13. J. Maldacena, Adv. Theor. Math. Phys. ${\bf 2:231}$ (1998),
hep-th/9711200.

\end{document}